\documentclass[aps, prb, reprint, longbibliography, floatfix, superscriptaddress, 10pt]{revtex4-2}
\usepackage[utf8]{inputenc}
\usepackage{amsmath}
\usepackage{amssymb}
\usepackage{times}
\usepackage[american]{babel}
\usepackage{csquotes}
\MakeOuterQuote{"}
\usepackage{bm}
\usepackage{graphicx}
\usepackage{xcolor}
\usepackage[caption=false]{subfig}
\usepackage{hyperref}
\hypersetup{colorlinks, citecolor=blue, linkcolor=violet, urlcolor=blue}
\usepackage{tikz}
\usetikzlibrary{shapes}

\newcommand{\uv}{\mathbf{e}}
\newcommand{\rr}{\mathbf{r}}
\newcommand{\irm}{\mathrm{i}}
\newcommand{\erm}{\mathrm{e}}
\newcommand{\drm}{\mathrm{d}}

\DeclareSymbolFont{rmlargesymbols}{U}{euex}{m}{n}
\DeclareMathSymbol{\rmintop}{\mathop}{rmlargesymbols}{82}
\newcommand{\rmint}{\rmintop\nolimits}

\graphicspath{{Figs/}}
\begin{document}
\title{Dynamics of a pair of magnetic dipoles with non-reciprocal interactions due to a moving conductor}

\author{Artem Rakcheev}
\email{artem.rakcheev@psi.ch}
\affiliation{Laboratory for Theoretical and Computational Physics, Paul Scherrer Institut, 5232 Villigen PSI, Switzerland}
\affiliation{Institut für Theoretische Physik, Universität Innsbruck, A-6020 Innsbruck, Austria}

\author{Andreas M. Läuchli}
\affiliation{Laboratory for Theoretical and Computational Physics, Paul Scherrer Institut, 5232 Villigen PSI, Switzerland}
\affiliation{Institute of Physics, Ecole Polytechnique Fédérale de Lausanne (EPFL), CH-1015 Lausanne, Switzerland}
\affiliation{Institut für Theoretische Physik, Universität Innsbruck, A-6020 Innsbruck, Austria}

\date{\today}

\begin{abstract}
Recently it was demonstrated theoretically and experimentally, that the presence of a moving conductor can break the reciprocity in the interactions between magnetic dipoles. In this article we investigate the influence of non-reciprocity on the dynamics of a pair of rigid XY dipoles, which have been realized in experiments. In particular, we focus on the energy non-conservation, which is a consequence of the non-reciprocity. We find that the dynamics indeed has regimes, wherein the kinetic energy grows quadratically. However, whether energy absorption occurs, depends strongly on the initial conditions on the dipoles. Simulations for various initial conditions reveal an intricate dependence, resulting in a rich structure of the energy absorbing regime in the initial condition space. Nevertheless, we provide a qualitative explanation of these observations, interpreting the absence of energy absorption as a confinement of the dynamics in phase space. 
\end{abstract}
\maketitle

\section{\label{sec: intro} Introduction}
It has been well known for centuries, that currents are induced in a conductor moving through a (static) magnetic field~\cite{raith_experimentalphysik_2006}. These, so-called \textit{eddy currents}, can lead to heating of the conductor - effectively transforming kinetic into thermal energy. This effect has been used in industrial applications, for instance magnetic brakes for decades~\cite{raith_experimentalphysik_2006}, with new applications such as velocimetry being proposed/implemented more recently~\cite{thess_lorentz_2006}. Very recently, the effects on the interactions between multiple dipoles close to a moving conductor have been investigated theoretically and experimentally~\cite{prat-camps_circumventing_2018}. In this work it was demonstrated for the first time, that in setups with a pair of dipoles and a moving conductor magnetic reciprocity can be broken.

The term \emph{non-reciprocity} is used rather broadly in the literature, typically referring to some sort of inequivalence of interactions between different parts of a system, and we will define the exact conditions for magnetic non-reciprocity later. A more well-defined, and closely related, concept is the breaking of Newton's third law \textit{"actio est reactio"}. Since all fundamental interactions are reciprocal, non-reciprocity can only arise in an effective description. In our case for example, we will focus on the dipolar degrees of freedom and treat the conductor as an environment, whose sole effect is to modify the interactions. Systems in which non-reciprocity and/or the breaking of Newton's third law have been discussed, range from particles in a plasma~\cite{lisin_experimental_2020} and acoustic surface waves~\cite{rasmussen_acoustic_2021} to robotic metamaterials~\cite{brandenbourger_nonreciprocal_2019}. Overviews and reviews of different phenomena can be found in~\cite{ivlev_statistical_2015, kryuchkov_dissipative_2018, fruchart_nonreciprocal_2021} and in~\cite{caloz_electromagnetic_2018, asadchy_tutorial_2020}, with the latter focusing on electromagnetic phenomena.

In~\cite{ivlev_statistical_2015, fruchart_nonreciprocal_2021} it was shown, that the framework of statistical physics can be extended to incorporate non-reciprocal systems. In particular, linear non-reciprocal systems, whose dynamics can be described by a non-Hermitian matrix, \textit{exceptional points} can be used to define phases and investigate phenomena such as synchronization of rotors. Fundamental differences to Hamiltonian (Hermitian) systems exist however - for example due to energy non-conservation, the system can absorb energy from the environment~\cite{kryuchkov_dissipative_2018}. In this article we will focus on this effect for a pair of dipoles. 

In Sec.~\ref{sec: setup} we will describe the setup in which magnetic reciprocity is broken and give a precise definition of reciprocity for magnetic dipoles in terms of the coupling matrix. We then briefly analyze the coupling matrix in our setup in Sec.~\ref{sec: couplings}, describe how to compute the coupling matrix numerically and show some numerical results supporting reciprocity breaking. In Sec.~\ref{sec: dynamics}, the main part of the article, we analyze the dynamics of a pair of rigid XY dipoles motivated by the experimental setup in~\cite{mellado_macroscopic_2012, concha_designing_2018, cisternas_stable_2021}. We derive the equations of motion, which turn out to be a system of coupled non-linear equations, and solve these numerically. Here, we will demonstrate that there can be a dynamical regime, wherein the pair absorbs (kinetic) energy, but also show that the occurrence of absorption is strongly dependent on the initial condition and that there are also non-absorbing regimes. Finally, we offer a qualitative explanation for the existence of both regimes, by analyzing the dynamics in phase space. This analysis suggests that the regimes are related to confinement of the dynamics in phase space. 
\section{\label{sec: setup} Setup and non-reciprocity}
\begin{figure}[htbp]
\centering
\includegraphics[width=0.98\linewidth]{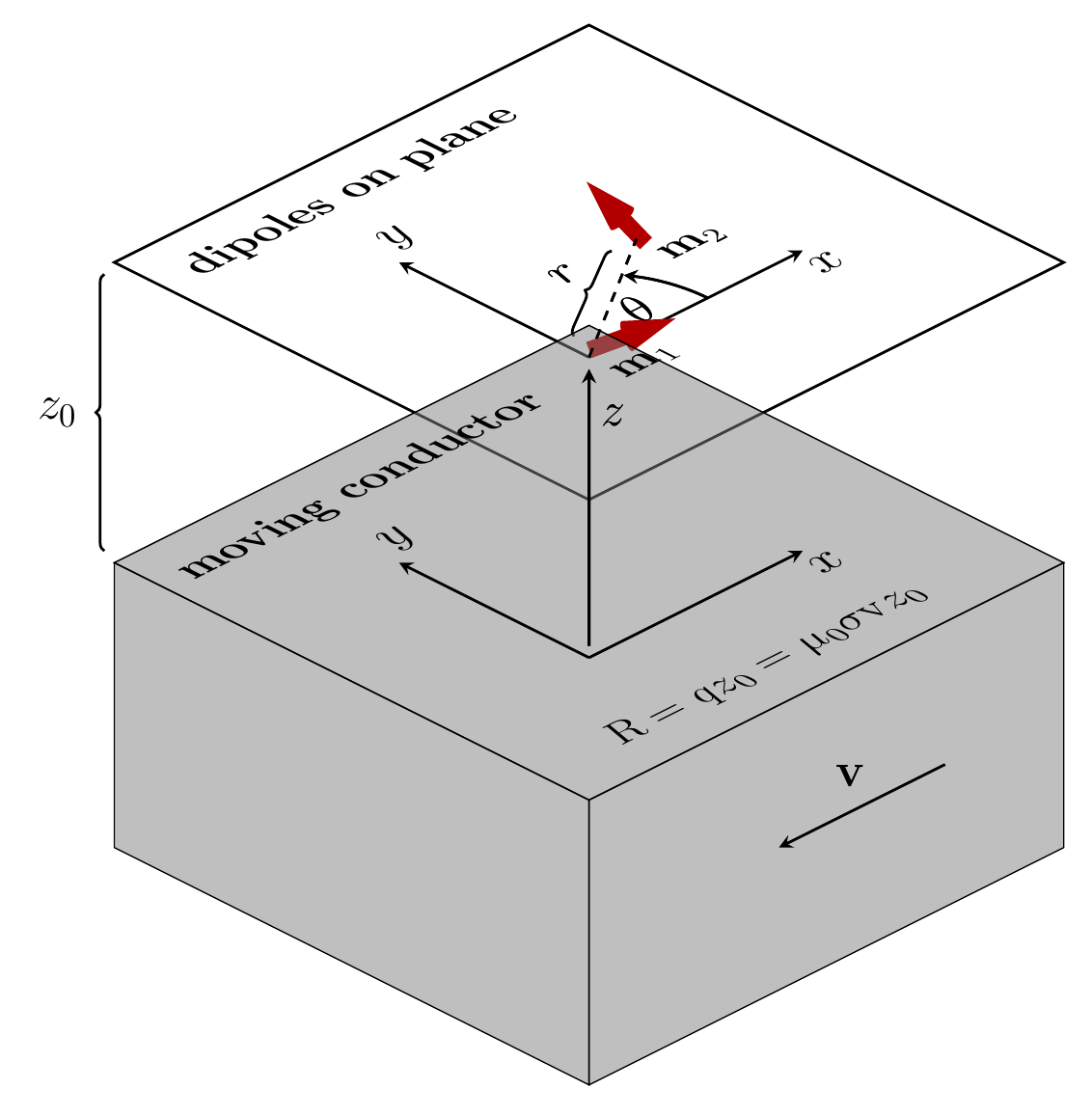} 
\caption{\label{fig: sketch of geom} Geometry of the system: A conductor with static conductivity $\sigma$ covers the lower half-space and is moving at a constant velocity along the negative $x$-axis. Magnetic dipoles with three-dimensional magnetic moments $ \mathbf{m}_{i} $ of constant magnitude are in a coplanar configuration at a distance $ z=z_{0} $ with respect to the interface. The interaction of a pair of dipoles depends on their relative position expressed by the distance $ r $ and the angle $ \theta $.}
\end{figure}

In~\cite{prat-camps_circumventing_2018} it was shown, that a setup consisting of a semi-infinite conductor moving at constant velocity and magnetic dipoles in a parallel plane, leads to non-reciprocal interactions between the dipoles. We follow this approach to reciprocity breaking by treating magnetic dipoles with magnetic moments $\mathbf{m}_i$ of constant magnitude $m$, located at a distance $z_{0} > 0$ above a non-magnetic (relative magnetic permeability $\mu=1$) conductor with static conductivity $\sigma$ that moves at a constant velocity $v<0$, with respect to the dipoles, along the $x$-axis. The conductor extends over a half-space with an interface that coincides with the $xy$-plane at $z=0$. The entire setup is sketched in Fig.~\ref{fig: sketch of geom}. This model neglects transmissions and reflections from the finite geometry of the conductor, which we expect to be negligible for conductors with a sufficiently large static conductivity~\cite{buhmann_dispersion_2012, prat-camps_circumventing_2018}. Furthermore, we will neglect any feedback on the conductor, which could lead to kinetic energy losses and subsequent slowing down, including losses due to the aforementioned \textit{eddy currents}, as well as losses due to the energy absorption to be discussed later.

%%%%%%%%%%%%%%%%%%%%%%%%
%%%%%%%%%%%%%%%%%%%%%%%%

\subsection{Single-dipole field}
First, imagine a single dipole above the conductor. Due to the motion, the conductor experiences a changing magnetic field, leading to the induction of eddy currents, which in turn create an additional scattered magnetic field~\cite{ thess_lorentz_2006, thess_theory_2007, raith_experimentalphysik_2006}. The total field is then given by the sum of the original and the scattered field and can be described by the coupling matrix $\mathbb{K}$
\begin{equation}
\label{eq: field from coupling matrix}
\mathbf{B}(\mathbf{r})=\mathbb{K}(\mathbf{r})\mathbf{m},
\end{equation} 
where we assume the dipole to be located at the origin. The original field of a dipole, also leading to the reciprocal dipole-dipole interaction, corresponds to the coupling matrix~\cite{jackson_classical_1998, coey_magnetism_2010}
\begin{equation}
\label{eq: coupling matrix dipole-dipole}
\mathbb{K}_{\mathrm{dpl}}(\mathbf{r})=\frac{\mu_{0}}{4\pi}\frac{3\hat{\mathbf{r}} \otimes \hat{\mathbf{r}}-\mathbb{I}}{r^{3}},
\end{equation}
where $r=|\mathbf{r}|$, $\hat{\mathbf{r}}=\mathbf{r}/r$, $\mathbb{I}$ is the identity matrix and $\otimes$ denotes the outer product of two vectors.

The scattered field in similar setups has been investigated using different analytic and numeric techniques~\cite{thess_lorentz_2006, thess_theory_2007, votyakov_interaction_2012, prat-camps_circumventing_2018}. We will follow the derivation presented in~\cite{prat-camps_circumventing_2018}, which provides the coupling matrix for an arbitrary orientation of the dipole and velocity of the conductor, although in this work we focus on non-relativistic velocities and hence the results correspond to the first terms in the appropriate expansion. Hereafter, we outline the important steps of the derivation, which can be found in more detail in Appendix~\ref{app: Derivation of the Potential Energy}. We model the linear electrodynamic response of the conductor by an isotropic and complex relative permittivity $\epsilon(\omega)= 1+\irm \sigma/(\epsilon_0\omega)$~\cite{ashcroft_solid_1976}, where $\epsilon_0$ denotes the vacuum permeability. In conjunction with the dyadic Green's function for a conducting half-space~\cite{buhmann_dispersion_2012}, we can express the total electromagnetic field as a function of the source, which, in the rest-frame of the moving conductor, consist of a dipole with electric and magnetic dipole moment. This allows us to derive an analytical expression for the magnetic (and electric) field emanating from the dipole, including the contribution of the induced currents $\mathbb{K}_{\mathrm{ind}}$. In the non-relativistic limit, the contributions from the electric terms become vanishingly small and can safely be neglected. Therefore, the interaction between multiple dipoles is dominated by the magnetic field. We will discuss the result for $\mathbb{K}_{\mathrm{ind}}$ in detail in Sec.~\ref{sec: couplings}, but let us first mention some general aspects of magnetic non-reciprocity.

 \subsection{Magnetic non-reciprocity}
To begin the discussion we note, that due to the linearity of the Maxwell equations, the induced currents in case of multiple dipoles are a superposition of the single dipole case and the same holds for the scattered fields. Therefore, the field of each individual dipole can still be described by Eq.~\eqref{eq: field from coupling matrix} using the single dipole coupling matrix, with $\mathbf{r}$ shifted appropriately~\footnote{A coordinate shift does not change $\mathbb{K}$ due to translational invariance.}. In~\cite{prat-camps_circumventing_2018} it was argued, that for dipoles the magnetic reciprocity condition is broken, if $\mathbf{m}_{1}\mathbf{B}_{2}\neq \mathbf{m}_{2}\mathbf{B}_{1}$. Here, the fields $\mathbf{B}_{i}$ denote the field created by the $i$-th dipole. This condition has also been experimentally tested in~\cite{prat-camps_circumventing_2018}, by measuring the fields for some selected orientations of the dipoles. 

The condition has a simple interpretation, in view of the expression for the energy of a dipole in a field~\cite{jackson_classical_1998, kholmetskii_electric_2014}
 \begin{equation}
 \label{eq: energy of dipole in field}
  E=-\mathbf{m}\mathbf{B}.
 \end{equation}
 As one can see, breaking of the reciprocity condition occurs, if the energy is different, depending on which dipole is taken as the source of the field. If both energies do not coincide, the dynamics is not governed by a (global) Hamiltonian. Furthermore, Newton's third law is broken, since the force $\nabla(\mathbf{m}\mathbf{B})$ on each dipole is not equal and opposite in general, if the expressions do not coincide. By inserting Eq.~\eqref{eq: field from coupling matrix}, we obtain the reciprocity condition for the coupling matrix~\footnote{As discussed later, there is also a self-interaction, which does not change the argument though.}
 \begin{align}
  \label{eq: hamiltonian of pair condition}
  -\mathbf{m}_{2}\mathbb{K}(\mathbf{r})\mathbf{m}_{1} &\overset{!}{=} -\mathbf{m}_{1}\mathbb{K}(-\mathbf{r})\mathbf{m}_{2} \nonumber\\
 \Leftrightarrow  \mathbf{m}_{1}\mathbb{K}^{T}(\mathbf{r})\mathbf{m}_{2} &\overset{!}{=} \mathbf{m}_{1}\mathbb{K}(-\mathbf{r})\mathbf{m}_{2} \nonumber\\
 \Rightarrow  \mathbb{K}^{T}(\mathbf{r}) &\overset{!}{=} \mathbb{K}(-\mathbf{r}).
 \end{align}
As we will see in Sec.~\ref{sec: couplings}, the equality is generally violated in our setup. 

\section{\label{sec: couplings} Coupling matrix}
As shown in Appendix~\ref{app: Derivation of the Potential Energy}, the induced coupling matrix can be formally expressed using a double integral
\begin{equation}
\label{eq: coupling matrix integral}
\mathbb{K}_{\mathrm{ind}}(\mathbf{r})=\frac{\mu_{0}}{8\pi^2 z^{3}_{0}}\rmint\limits_{0}^{\infty}\drm\xi \; \erm^{-2 \xi} \xi^{2} \rmint\limits_{0}^{2\pi} \drm \phi \;  r_{s}(\xi, \phi)\erm^{\irm \frac{\xi}{z_{0}}\mathbf{r} \cdot \mathbf{e}_{\rho}}\mathbb{M}(\phi),
\end{equation}
where $r_{s}$ denotes reflection coefficient of an infinite conducting half-space \cite{buhmann_dispersion_2012}
\begin{equation}
\label{eq: scattering function}
r_{s}(\xi, \phi)=\frac{\xi-\sqrt{\xi^{2}-\irm R\xi\cos(\phi)}}{\xi+\sqrt{\xi^{2}-\irm R\xi\cos(\phi)}}
\end{equation}
and the matrix in the integrand is
\begin{equation}
\mathbb{M}(\phi)=(\mathbf{e}_{\rho}+\irm\mathbf{e}_{z})\otimes(\mathbf{e}_{\rho}-\irm\mathbf{e}_{z})
\end{equation}
with $\mathbf{e}_{\rho}=(\cos(\phi), \sin(\phi), 0)^{T} ,\; \mathbf{e}_{z}=(0, 0, 1)^{T}$. We analyze the integral expression in detail in Appendix~\ref{app: coupling details} and constrain the discussion to the main results in the following. 

First, we note that we can decompose the full coupling matrix (including the dipole-dipole interactions) into four terms, corresponding to different parities upon (spatial) inversion ($\mathbb{K}(\mathbf{r}) \to \mathbb{K}(-\mathbf{r})$) and transposition ($\mathbb{K}(\mathbf{r}) \to \mathbb{K}^{T}(\mathbf{r})$). Only two of these fulfill the reciprocity condition stated in Eq.~\eqref{eq: hamiltonian of pair condition} and since all four contribute in general, the total coupling matrix is non-reciprocal. In the following, we will denote the terms with even/odd parity under inversion by $\mathbb{K}^{+/-}$. 

The reciprocal terms can be identified as~\cite{dzyaloshinsky_thermodynamic_1958, moriya_anisotropic_1960, chikazumi_physics_2009, blugel_computing_2014, li_spin_2021} a (symmetric) exchange term resulting from the reciprocal part of $\mathbb{K}^{+}$, an antisymmetric exchange (Dzyaloshinsky-Moriya) term resulting from the reciprocal part of $\mathbb{K}^{-}$ and a single-ion anisotropy corresponding to the limit $\lim\limits_{r \to 0}\mathbb{K}^{+}$. In general the non-reciprocal parts do not vanish and can be of comparable magnitude as the reciprocal ones. However, in case of the "perfect conductor" ($\sigma \to \infty$) they do \footnote{Of course they also vanish in the trivial case of the stationary conductor $v=0$}. Furthermore, in this limit the integrals can be evaluated analytically as shown in Appendix~\ref{app: perfect cond}. A notable result from this calculation, is that the exchange terms decay as $r^{-3}$ and the Dzyaloshinsky-Moriya terms as $r^{-4}$ (in this limit), in agreement with the method of images~\cite{jackson_classical_1998}.

\subsection{Numerical Evaluation}
Except for this limiting case, one needs to evaluate the integrals using numerical methods.
If the variables in the coupling integrals are measured appropriately ($r\to r/a,\; z_{0} \to z_{0}/a, \; q \to qa$) the resulting matrix $\mathbb{K}$ is dimensionless. For the computations we set dimensionless units by introducing a length scale $a$ and by using $\mu_{0}/32\pi^{2}a^{3}$ as a base unit for the couplings. In Appendix~\ref{app: experiment} we discuss the real values of these parameters in a possible experimental setup. 

After appropriate variable transformations (see Appendix~\ref{app: coupling details} for details), the angular and radial integrals are of a form suitable in principle for Gauss-Chebyshev (measure $ \sqrt{1-u^{2}}^{-1} $) and generalized Gauss-Laguerre Quadrature (measure $ x^{2}\erm^{-x} $) respectively~\cite{press_numerical_2007}. These methods approximate an integral by a sum 
\begin{equation*}
    \int\limits_{a}^{b} f(x) \; \drm x \approx \sum\limits_{i=1}^{n} w_{i}f(x_{i}),
\end{equation*}
where $w_{i}$ are the \emph{weights}, $x_{i}$ the \emph{nodes}, and $n$ the \emph{order} of the quadrature. A quadrature of order $n$ is \emph{exact} for integrating polynomials up to degree $2n -1$ multiplied by the measure, if the nodes and weights are chosen correctly (in our case we obtain them using inbuilt \emph{SciPy} routines). The integration boundaries $a, b$ vary based on the measure; the relevant boundaries in our case are shown in Eq.~\ref{eq: app couplings units}. However, while Gauss-Chebyshev quadrature can indeed be used efficiently for the angular integral, the weights for generalized Gauss-Laguerre Quadrature start to be limited by numerical precision at an order of around 200, which is not enough to resolve the high spatial frequency in the exponential $\exp\left(\irm \frac{\xi}{z_{0}}\mathbf{r} \cdot \mathbf{e}_{\rho}\right)$ for many parameter choices. Therefore, the radial integral needs other methods, and we settled for an adaptive integration approach, to be discussed in the following. The single-ion anisotropy terms however do not have this oscillatory behavior and therefore can be evaluated using a Gauss-Laguerre Quadrature.

\subsubsection*{Radial Integral}
Due to the scattering function having a singularity at its derivative for $ x \rightarrow 0 $ (and the further complication due to the semi-infinite domain), usual methods for strongly oscillatory integrals such as Levin or Filon type approaches \cite{olver_numerical_2008} are not applicable to our best understanding. Thus, we settle for a brute force approach by truncating the integral at a finite $ x_{max} $ and taking enough points to resolve the oscillations. The truncation is based on the maximum of the measure $ \frac{2}{R} $. We find that taking a factor of 10 is sufficient in all regimes discussed in this work. Within this region though, one needs to resolve the oscillations with frequency $ \Omega $, which we accomplish by adaptively choosing the number of integration points, such that each period is resolved with at least one hundred points. To be precise, given the truncation and the frequency, the exact number of points is selected such that it is suitable for Romberg integration ($ 2^{n} + 1 $ points)~\cite{press_numerical_2007}. Furthermore, at least 8193 points are used by default, irrespective of the parameters. 
\subsubsection*{Angular Integral}
As discussed above the angular integral can be evaluated using Gauss-Chebyshev Quadrature. The necessary order depends on the parameters and on the distance $ r $, up to which the couplings are to be evaluated. Generally the necessary order increases with $ r, q, \frac{1}{z_{0}} $ and needs to be obtained from manual convergence analysis, whereby the couplings are evaluated with different orders for some fixed angles $ \theta $. The order for all computations in this article is 10000, which we found to be sufficient for the range of parameters.
\subsection{Visualization}
\begin{figure*}[htbp]
\subfloat[$qa=0.1$]{\label{fig: couplings polar qa=0.1} \includegraphics[width=0.48\textwidth]{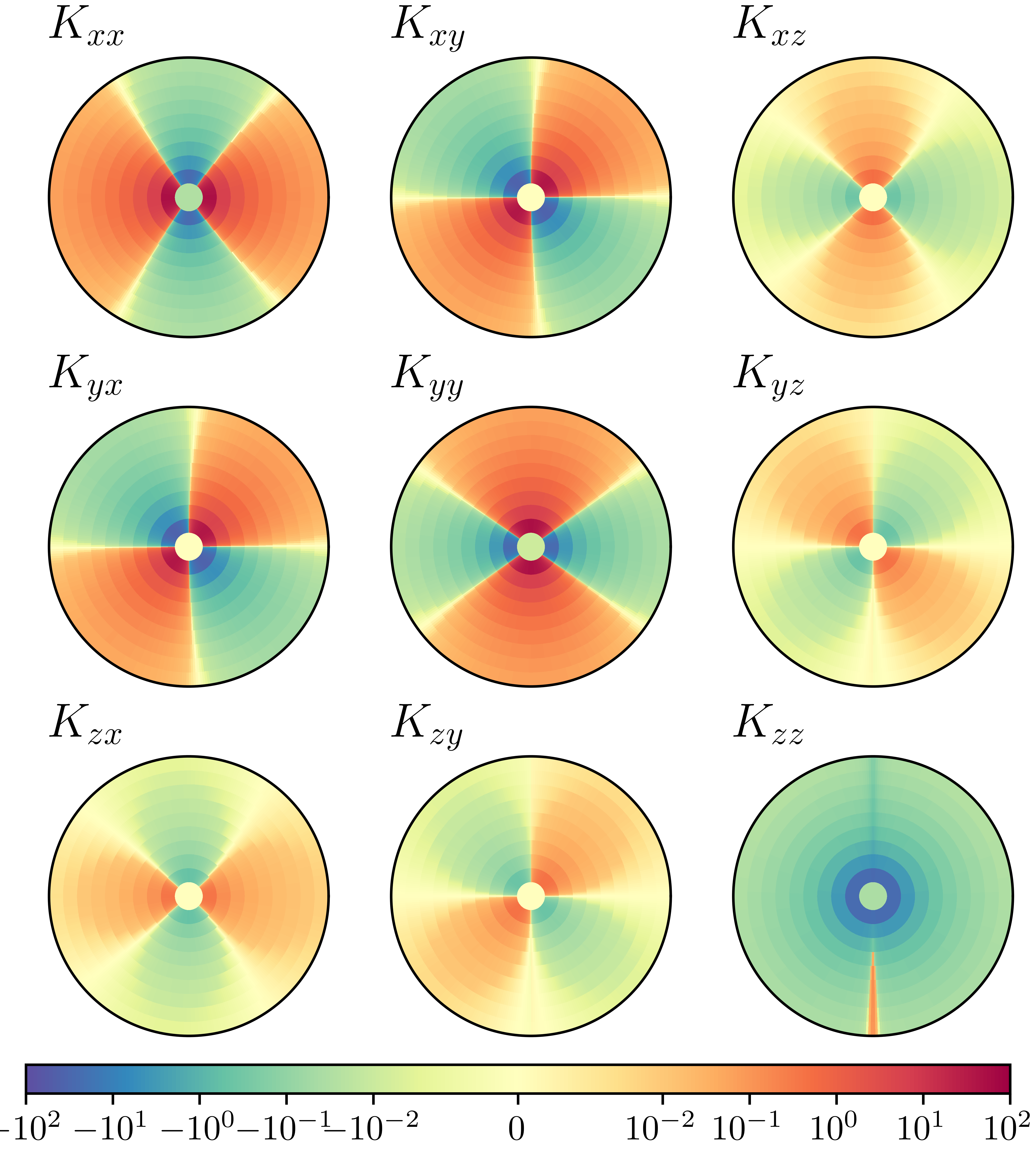}}
\subfloat[$qa=10$]{\label{fig: couplings polar qa=10} \includegraphics[width=0.48\textwidth]{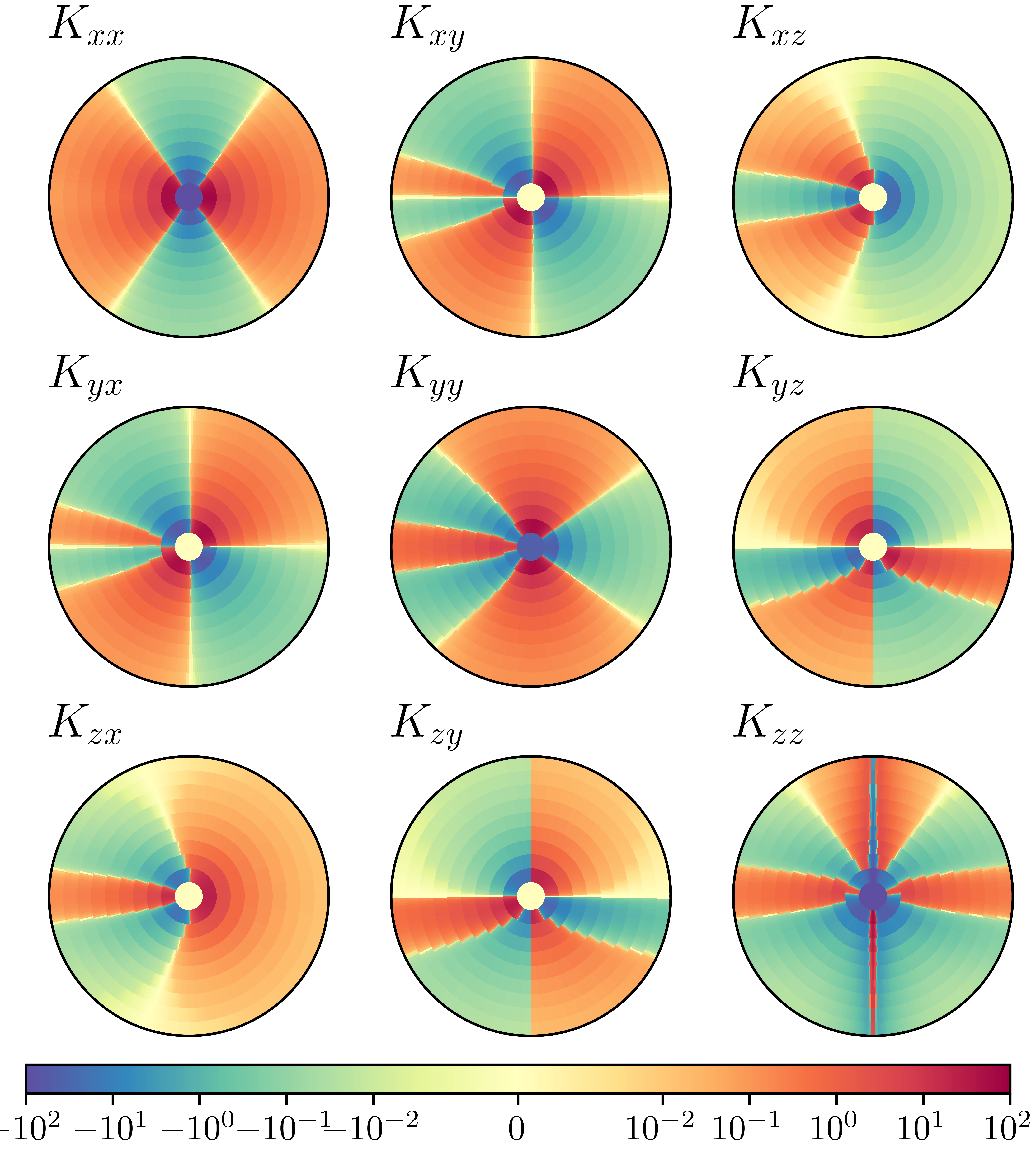}}
\caption{\label{fig: couplings_polar} Full couplings for radii $r \leq 10$ with $z_{0} / a=0.1$ and $qa=0.1$ (a) and $qa=10$ (b). In (a) the dipole-dipole interactions dominate, while in (b) the induced interactions are strongest. The major differences are the induced $xz$- and $yz$-couplings and the axial features around the x-axis. Reciprocity is visibly broken in (b) - for instance $K_{yy}(-\mathbf{r})\neq K_{yy}(\mathbf{r})$.}
\end{figure*}

The spatial dependence of the full interactions can be visualized using polar plots as in Fig.~\ref{fig: couplings_polar}. Here we see the full coupling matrix at a small $(qa=0.1)$ and a large $(qa=10)$ value of $qa$. The couplings are labeled by the components of the dipole moments that they would couple through the interaction $-\mathbf{m_{2}}\mathbb{K}\mathbf{m}_{1}$ - for example $K_{xy}$ is the coefficient of $m^{x}_{2}m^{y}_{1}$ in the resulting sum. In the plots, the couplings are displayed as a function of distance and the angle between dipoles (with $x$ and $y$ axes defined as in Fig.~\ref{fig: sketch of geom}). At $qa=0.1$, the dipole-dipole interactions dominate with respect to the induced terms. Since these only include the couplings $K_{xx}, K_{xy}=K_{yx}, K_{yy}$ and $K_{zz}$, the others are barely noticeable even on a logarithmic (color) scale. The total couplings are also nearly reciprocal, but first slight deviations are visible. At $qa=10$, the other couplings are clearly visible and of comparable strength at some angles. Furthermore, most of the symmetric exchange terms are significantly modified, featuring strong axial features. The inversion $\mathbf{r} \to -\mathbf{r}$ corresponds to a half-turn ($\theta \to \theta + \pi$) and the transposition $\mathbb{K}^{T}$ to an exchange of component indices $K_{ab} \to K_{ba}$. Therefore, the breaking of reciprocity based on the condition from Eq.~\eqref{eq: hamiltonian of pair condition} is clearly visible at $qa=10$; for example in $K_{yy}$.
%%%%%%%%%%%%%%%%%%%%%%%
%%%%%%%%%%%%%%%%%%%%%%%%
\section{\label{sec: dynamics} Dynamics of rigid planar dipoles}
Having seen how dipoles interact with each other in the presence of the conductor, we now investigate the consequences of non-reciprocity for the dynamics of a pair. In various experimental setups~\cite{mellado_macroscopic_2012, arnalds_thermal_2014,leo_collective_2018,concha_designing_2018,cisternas_stable_2021}, two-dimensional magnetic moments (XY rotors) can be realized. Especially, given that the scales in the setup of~\cite{mellado_macroscopic_2012, concha_designing_2018,cisternas_stable_2021} could be suitable to observe the effects of non-reciprocity, as discussed in Appendix~\ref{app: experiment}, we will look into this type of setup more closely in the following. The main features are rigid body dipoles (rods in the experiment) i.e. the magnetic moment is "locked" to a spatial orientation of the rigid body and the constraint of the moments to two dimensions (here the $xy$-plane). Both of these factors influence the form of the equation(s) of motion (EOM), as we will see below, but apart from the form of the EOM, the following treatment does not use further experimental details.
\subsection{Equation of motion}
The EOM for a single dipole, can be derived from the torque on a magnetic dipole in a field~\cite{jackson_classical_1998, kholmetskii_electric_2014}
\begin{equation}
\mathbf{T}=\mathbf{m}\times \mathbf{B}
\end{equation}
and the dynamical equation for a rigid body~\cite{luders_experimentalphysik_2008} rotating around the $z$-axis
\begin{equation}
T_{z}=I\ddot{\varphi},
\end{equation}
with the moment of inertia $I$ and the angle $\varphi$ representing the orientation of the dipole relative to the $x$-axis.
Introducing a further friction term appearing in the experiments~\cite{mellado_macroscopic_2012, concha_designing_2018, cisternas_stable_2021}, the EOM reads
\begin{equation}
I\ddot{\varphi}=\left(\mathbf{m}\times \mathbf{B}\right)\cdot \mathbf{e}_{z}-\eta \dot{\varphi}.
\end{equation}
An alternative derivation using the Euler-Lagrange equation is presented in Appendix~\ref{app: eom}.

Note that here we do not use the standard formula for Larmor precession~\cite{coey_magnetism_2010} $\dot{\mathbf{m}}=\mathbf{T}$, because that assumes the magnetic moment to be directly related to the angular momentum $\mathbf{m}\propto \mathbf{L}$, which is for instance the case if the magnetic moment results from an electron in orbit. In case of the rigid body however, the angular momentum stems from the rigid body rotation. 
\subsection{Pair of dipoles}
In the following section, we focus on the dynamics of a dipole pair and show that depending on the initial orientations of dipoles initially at rest (zero angular velocity), they can either be in an "oscillating" regime or in an (energy) "absorbing" regime, absorbing energy from the conductor in the latter. This energy absorption is a direct consequence from the non-reciprocity and the aforementioned inability to define a Hamiltonian, allowing energy non-conservation in the dipolar system. For the remainder of the section, we neglect the single-ion anisotropy, since a correct determination of the torque from this self-interaction would need to take into account the geometry of the rigid body. A discussion of the effect, showing that it does not change the main conclusions qualitatively, is presented in Appendix~\ref{app: dyn with sia}.

The dipoles are characterized by the angles $\varphi_{1/2}$ and $\mathbf{r}=\mathbf{r}_{2}-\mathbf{r}_{1}$. We set $r=1$, meaning that the length scale $a$ introduced in Sec.~\ref{sec: couplings} corresponds to the physical distance of the dipoles. Hence, $\mathbf{r}$ reduces to $\theta$, the angle relative to the $x$-axis. Again, we would like to choose units such that the EOM becomes dimensionless. For this, we introduce a timescale set by the system parameters $t_{\mathrm{sys}}=\sqrt{\frac{32 \pi^{2} I a^{3}}{\mu_{0}m^{2}}}$, where $m$ is the magnitude of the magnetic moment, and scale $t$ to $\tau=t/t_{\mathrm{sys}}$. In this way, all scales relevant for the couplings and for the description of the dipoles are captured by a single time scale. With damping we would have a second time scale $t_{\mathrm{damp}} = I / \eta$ and the dimensionless damping factor would be $t_{\mathrm{sys}}/t_{\mathrm{damp}}$. We discuss all relevant scales in view of the aforementioned experiments in Appendix~\ref{app: experiment}.

It turns out, that it is favorable to transform to the sum and difference variables $\varphi_{\pm}=\varphi_{1} \pm \varphi_{2}$ for which the EOM, expressing the field using the couplings, can be written as
\begin{widetext}
\begin{align}
\frac{d^{2}\varphi_{+}}{d\tau^{2}} &= K^{+}_{xy}(\mathbf{r})\cos(\varphi_{+})+\frac{K^{+}_{yy}(\mathbf{r})-K^{+}_{xx}(\mathbf{r})}{2}\sin(\varphi_{+})+\frac{K^{-}_{yy}(\mathbf{r})+K^{-}_{xx}(\mathbf{r})}{2}\sin(\varphi_{-}) \nonumber \\
\frac{d^{2}\varphi_{-}}{d\tau^{2}} &= -K^{-}_{xy}(\mathbf{r})\cos(\varphi_{+})+\frac{K^{-}_{xx}(\mathbf{r})-K^{-}_{yy}(\mathbf{r})}{2}\sin(\varphi_{+})-\frac{K^{+}_{yy}(\mathbf{r})+K^{+}_{xx}(\mathbf{r})}{2}\sin(\varphi_{-}).
\label{eq: angular acceleration pm}
\end{align}
\end{widetext}
At this point we can already make an important observation: in case of reciprocal interactions, the $K^{-}_{ab}$ terms vanish~\footnote{This follows since $\mathbb{K}^{T}=\mathbb{K}$, if only couplings of the $x$- and $y$-components are considered.}. Therefore, the equations \textit{decouple} i.e. $\ddot{\varphi}_{\pm}$ is only a function of $\varphi_{\pm}$. We note in passing, that the equation of $\ddot{\varphi}_{-}$ coincides with the EOM of a simple pendulum without the small angle approximation in that case. The non-reciprocal couplings also couple the sum and difference components, such that in general we are dealing with a system of second order coupled non-linear ODE's. To our best knowledge there are no analytical methods to solve such equations, therefore we resort to numerical methods. 

%%%%%%%%%%%%%%%%%%%%%%%%
\begin{figure*}[htbp]
    \includegraphics[width=\textwidth]{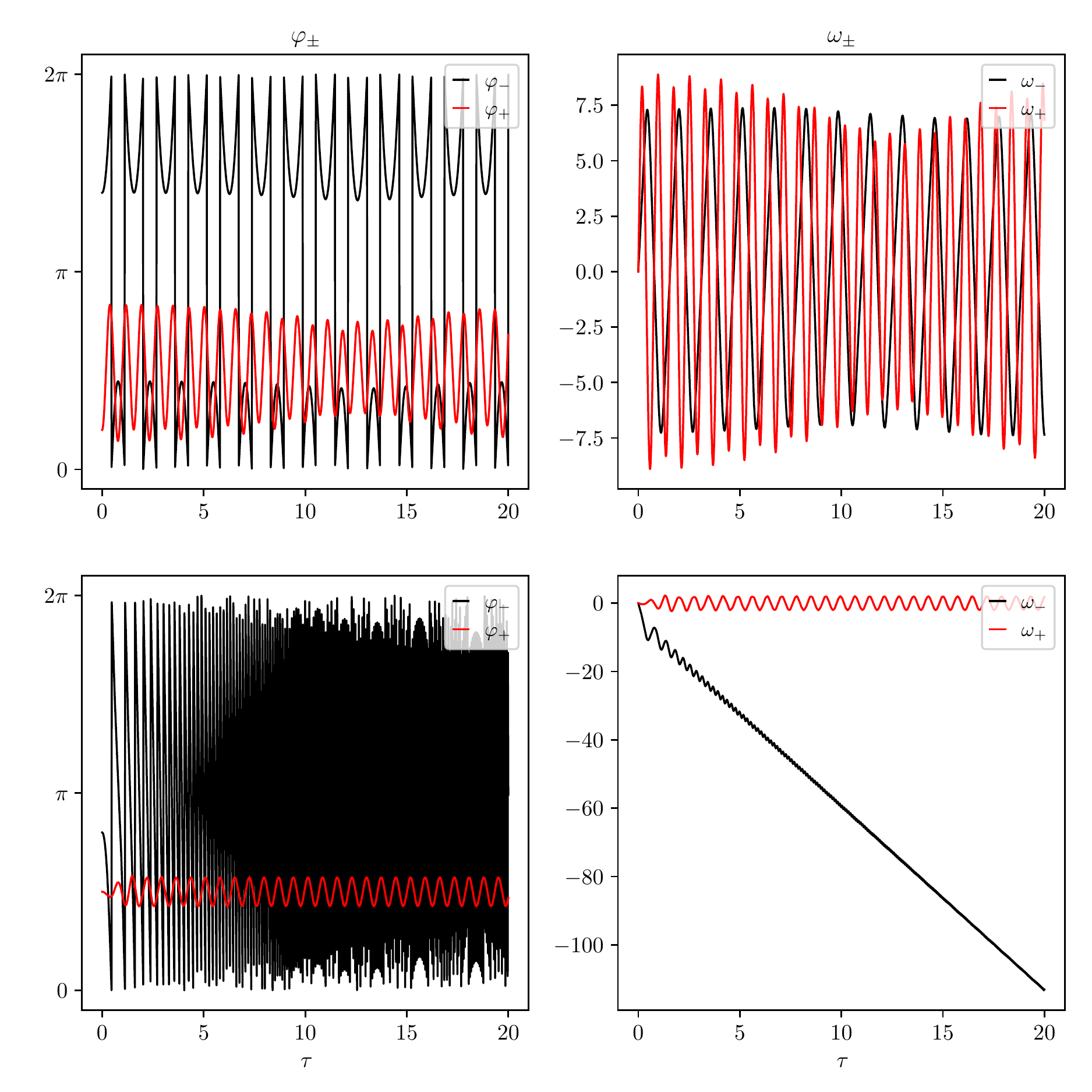}
    \caption{\label{fig: dipole dynamics} Dynamics of a pair of dipoles with $\theta=45^{\circ}$, $z_{0}/a=0.1$, $qa=1$. The plots show the evolution of the angular variables and the frequencies for two different initial conditions. The top plots correspond to oscillatory motion, while the bottom plots feature a linear growth of $\omega_{-}$.}
    \end{figure*}
%%%%%%%%%%%%%%%%%%%%%%%%

To be precise, we simulate the dynamics using \textit{SciPy} starting from various initial conditions for the angles $\varphi_{\pm}(0)$ and the dipoles being initially at rest $\omega_{\pm}(0)=0$, where $\omega$ denotes the angular frequency, and focus on the parameters $z_{0}/a=0.1$ and $0\leq qa \leq 1000$. For most orientations of the dipoles in space $\theta$ and values of $qa$ we can distinguish two dynamical regimes.
The different behavior can be seen in Fig.~\ref{fig: dipole dynamics}, wherein the dynamics of the angles $\dot{\varphi}_{\pm}$ and angular velocities $\dot{\omega}_{\pm}$ are shown for $\theta=45^{\circ}$ and $qa=1$. The plots show the dynamics for a simulation time of $\tau_{\mathrm{final}}=20$ starting from the initial conditions $\varphi_{+}(0)=0.65\pi,\; \varphi_{-}(0)=1.85\pi$ and $\varphi_{+}(0)=0.8\pi,\; \varphi_{-}(0)=1.4\pi$.
In the first case, the motion of all quantities is oscillatory around the initial values, while in the second a linear growth (modulo perturbations) of $\omega_{-}$ is observed, with $\omega_{+}$ oscillating around $0$.

The (linear) growth of $\omega_{-}$ also leads to a (quadratic) growth of the kinetic energy (density) $T=I(\omega^{2}_{+}+\omega^{2}_{-})/2L$. This non-conservation of energy is a further signature of non-reciprocal behavior, since, as we argued in Sec.~\ref{sec: setup}, there is no Hamiltonian to be conserved in the non-reciprocal case. As seen in Fig.~\ref{fig: dipole dynamics}, the contribution from $\omega_{-}$ dominates in case of a large energy absorption. In this case, the average angular acceleration (slope of the linear growth) can be approximated from the energy (density) by 
\begin{equation*}
    \ddot{\varphi}_{-}= \dot{\omega}_{-} \approx 2\sqrt{T/\tau_{\mathrm{final}}}.
\end{equation*}
In the following, we present the results of simulations of this quantity for varying initial conditions and try to get another point-of-view on the dynamics by looking at them in the $\varphi_{\pm}$ plane; we will refer to this plane as \emph{phase space} even though it is only a part of the full four dimensional phase space. 

%%%%%%%%%%%%%%%%%%%%%%%%
\begin{figure*}[htbp]
    \includegraphics[width=\textwidth]{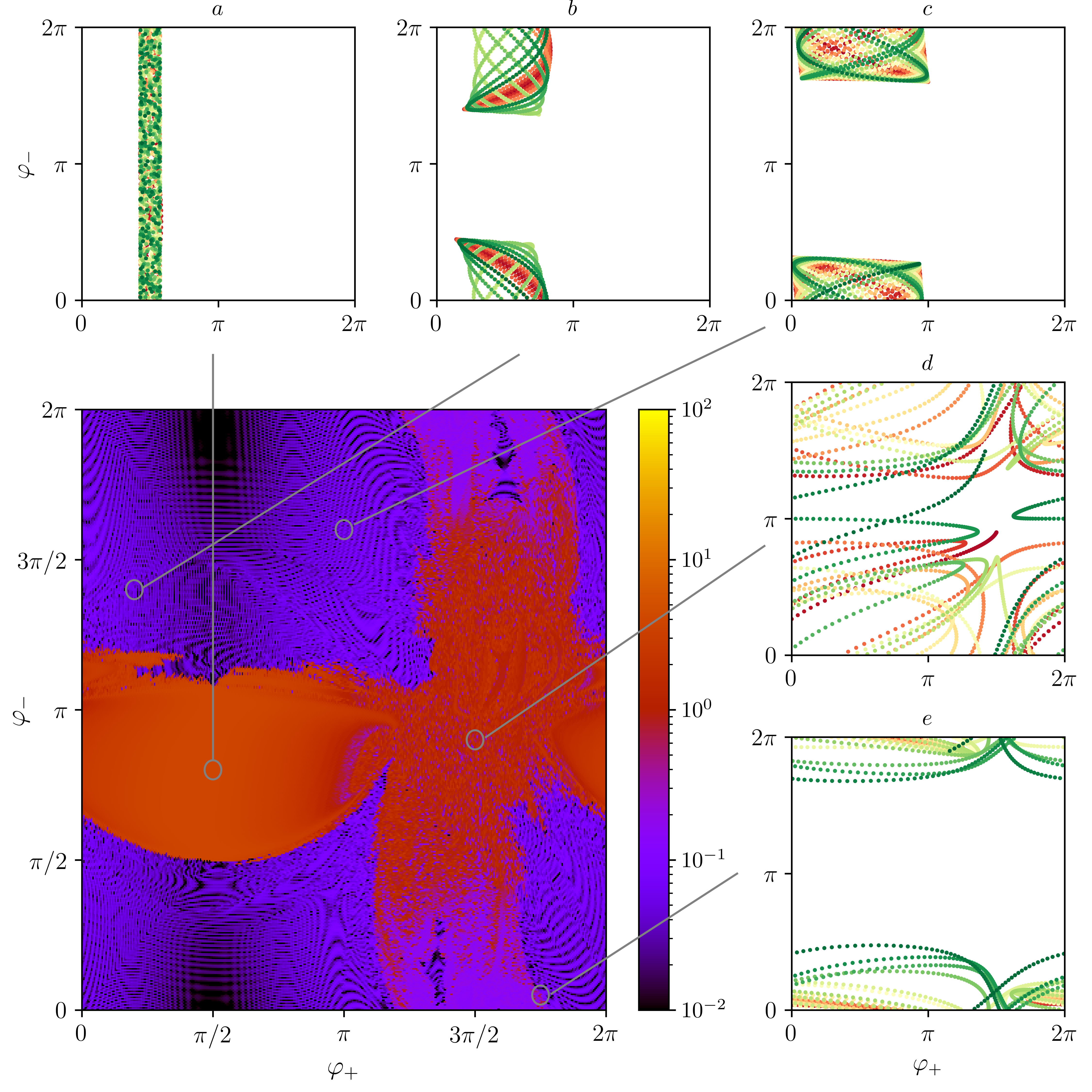}
    \caption{\label{fig: ear overview} Average angular acceleration and trajectories for $\theta=45^{\circ}$, $z_{0}/a=0.1$, and $qa=1$. In the main plot (bottom left) the average angular acceleration for $\tau_{\mathrm{final}}=100$ starting from different initial conditions is shown. The orange colored regions correspond to a large acceleration, while the purple / black regions correspond to essentially no acceleration (oscillatory motion). The smaller plots show trajectories for five select initial conditions (see Table~\ref{tab: initial conditions} for the exact values) and $\tau_{\mathrm{final}}=20$. The coloring indicates the evolution time going from red (start) to green (end). Here qualitatively different behavior is observed - see the main text for a detailed discussion and interpretation.}
    \end{figure*}
%%%%%%%%%%%%%%%%%%%%%%%%

All of this is shown in Fig.~\ref{fig: ear overview}: in the main plot on the bottom left we plot $\omega_{-}$ obtained from the energy density for various initial conditions and in the other subplots some trajectories for select initial conditions; all for an orientation of the dipoles of $\theta=45^{\circ}$, $z_{0}/a = 0.1$, $qa=1$, and $\tau_{\mathrm{final}}=100$. The select initial conditions labeled $a$ to $e$ are specified in Table~\ref{tab: initial conditions}.

\begin{table}
    \begin{tabular}{c | c c c c c}
        & $a$ & $b$ & $c$ & $d$ & $e$ \\
        \hline
        $ \varphi_{+}(0) $ & $0.5 \pi$ & $0.2 \pi$ & $1.0 \pi$ & $1.5 \pi$ & $1.75 \pi$ \\
        $ \varphi_{-}(0) $ & $0.8 \pi$ & $1.4 \pi$ & $1.6 \pi$ & $0.9 \pi$ & $0.05 \pi$ \\
    \end{tabular}
    \caption{\label{tab: initial conditions} Initial conditions for the trajectories in Fig.~\ref{fig: ear overview}.}
\end{table}

In the main plot we can see that the average acceleration ranges over several orders of magnitude and one can identify regions of high acceleration (orange) and low acceleration (purple/black). These regions show intricate features around the edges and also in their bulk for $\varphi_{+} \approx 3\pi/2$, which will be explained shortly. One should note though that some of the intricacies may result from averaging over a finite simulation time, which is probably responsible for the "stripy" pattern in the purple/black regions corresponding to oscillatory dynamics.

In the following we will build up a qualitative understanding for the different regimes by analyzing the phase space trajectories (visually). The trajectories show the evolution of $\varphi_{\pm}$ as a function of time. The color corresponds to the evolution time; red being the start and green the end of the simulation. Let us quickly summarize some observation in plots $a$ to $e$, before relating the trajectories to the energy gain. At this point one should also remember, that both angular parameters are circular and hence the parameter space is essentially a torus, leading  to the appearance of a "cutoff" in some trajectories.
\begin{enumerate}
    \item[$a\;$ -] The trajectory lies within a narrow strip along the $\varphi_{+}$ axis and traverse the entire $\varphi_{-}$ axis. For this trajectory the energy gain is particularly large.
    \item[$b\;$ -] The trajectory seems to lie within a slightly bent rectangular region, which is longer across the $\varphi_{-}$ axis. Here the energy gain is negligible.
    \item[$c\;$ -] Similar to $b$, but with a "straighter" and more "squareish" rectangle.
    \item[$d\;$ -] The trajectory seems to traverse the entire phase space. The energy gain seems reasonably high, but overall the initial condition lies in a region where the growth appears to be very sensitive to the initial conditions.
    \item[$e\;$ -] The trajectory is within a stripe as in $a$, however the stripe is now along the $\varphi_{+}$ axis, as opposed to the $\varphi_{-}$ axis. The energy growth is very low.
\end{enumerate}

%%%%%%%%%%%%%%%%%%%%%%%%
\begin{figure*}[htbp]
    \includegraphics[width=\textwidth]{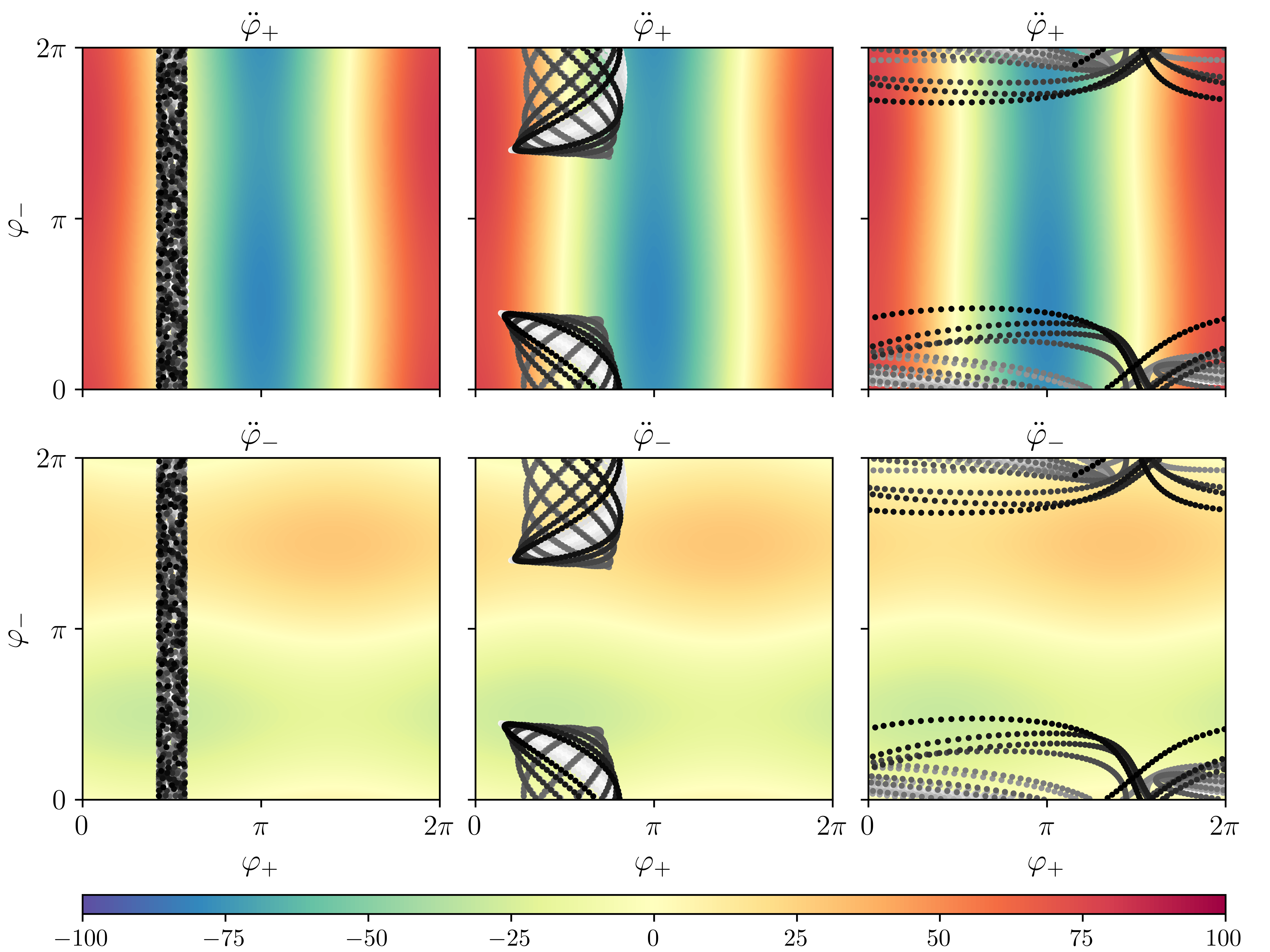}
    \caption{\label{fig: eom orbit} Dynamics in phase space based on the initial conditions $a$, $b$ and $e$ as in Fig.~\ref{fig: ear overview} and Table~\ref{tab: initial conditions}. The total evolution time is $\tau_{\mathrm{final}}=20$ with the passing of time indicated by color (white to black). In the background the accelerations $\ddot{\varphi}_{\pm}(\varphi_{+}, \varphi_{-})$ are plotted. In the left plot the motion is confined in $\varphi_{+}$ but traverses $\varphi_{-}$, corresponding to an energy growth. Then, in the second plot the dynamics is confined to a bend rectangle, corresponding to oscillatory motion. Finally, in the third plot the dynamics is confined in $\varphi_{-}$. This however does not correspond to a (significant) energy growth.} 
    \end{figure*}
%%%%%%%%%%%%%%%%%%%%%%%%

To understand this behavior, we take a second look at the trajectories $a$, $b$, and $e$ in Fig.~\ref{fig: eom orbit}; this time plotted on top of a background showing the acceleration function $\ddot{\varphi}_{\pm}(\varphi_{+}, \varphi_{-})$; with $\ddot{\varphi}_{+}$ at the top and $\ddot{\varphi}_{-}$ at the bottom. The trajectories are colored according to the time again, but this time going from white to black as time passes. Looking at the background, one can get a visual intuition for the qualitatively different dynamics. Here, we can make some observations explaining the appearance of energy growth and the general structure of the main plot in Fig.~\ref{fig: ear overview}. At first, we remember that in the reciprocal case Eqs.~\eqref{eq: angular acceleration pm} decouple, meaning that for example $\ddot{\varphi}_{+}$ is a function of $\varphi_{+}$ only. Visually this would mean that the background would have the same coloring along $\varphi_{-}$. In this case a Hamiltonian can be defined, and energy conservation defines boundaries in both directions. This would result in a bounding rectangle. Furthermore, we notice that in this case there will be a stable and an unstable fixed point in both $\ddot{\varphi}_{+}$ and $\ddot{\varphi}_{-}$. For $\ddot{\varphi}_{+}$ the stable one lies at $\ddot{\varphi}_{+}=\pi/2$ and the unstable one at $\ddot{\varphi}_{+}=3\pi/2$. For $\ddot{\varphi}_{-}$ they lie at $0$ and $\pi$. The final observation is that the coloring for $\ddot{\varphi}_{+}$ is much more saturated than its counterpart, indicating that the addition of non-reciprocal terms affects $\ddot{\varphi}_{-}$ more strongly.

%%%%%%%%%%%%%%%%%%%%%%%%
\begin{figure*}[htbp]
    \includegraphics[width=\textwidth]{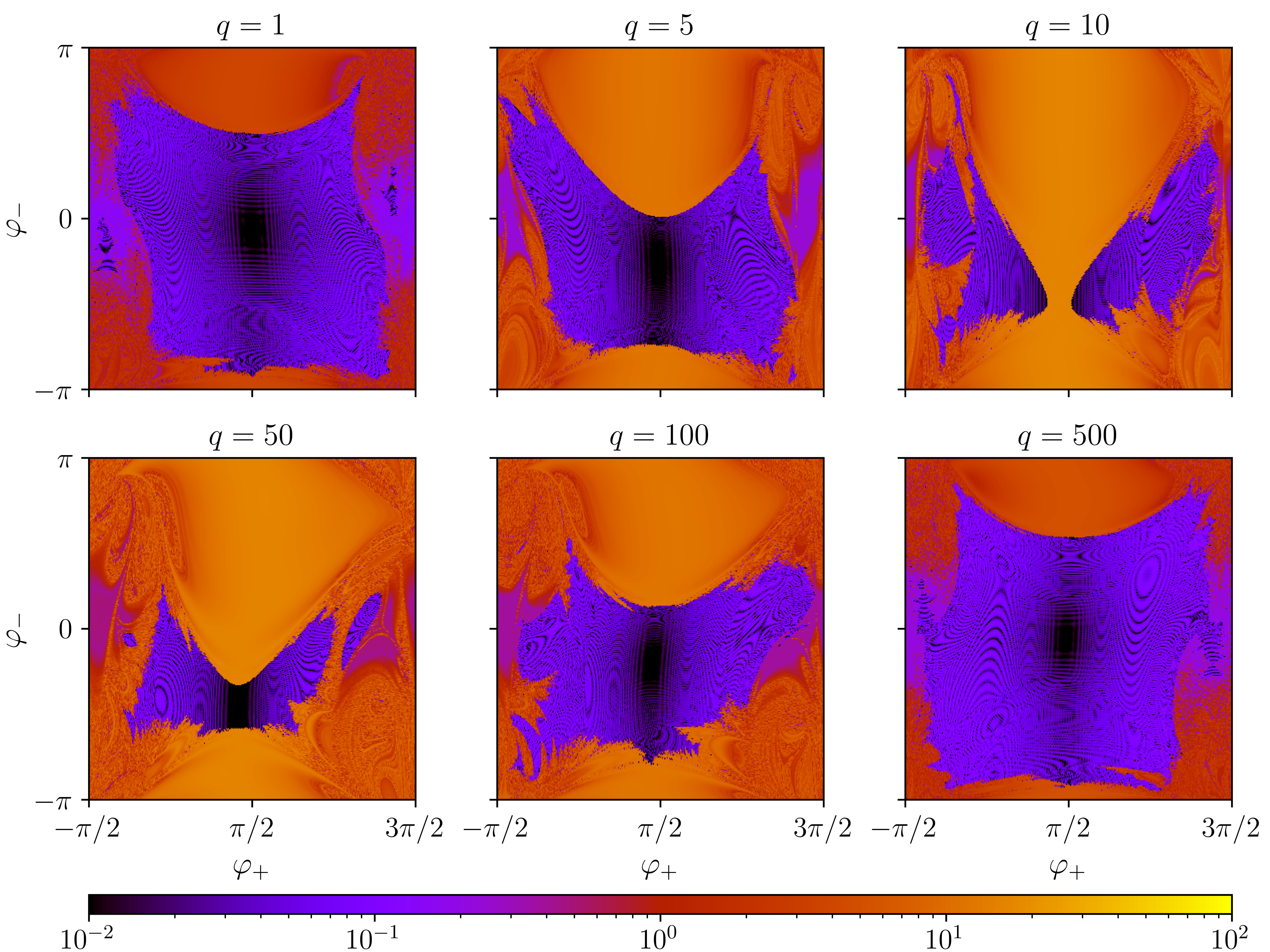}
    \caption{\label{fig: ear diagram} Averaged angular acceleration rate for $\theta=45^{\circ}$ and $z_{0} / a=0.1$ starting from different initial conditions. The total evolution time is $\tau_{\mathrm{final}}=100$. We observe orders of magnitude differences between different initial conditions. The absorbing regime is strongest at intermediate $qa$, since $qa=0, \; \infty$ both feature reciprocal couplings.}
    \end{figure*}
%%%%%%%%%%%%%%%%%%%%%%%%

With these ideas the interpretation of the dynamics is the following: in the case of energy growth the reciprocity breaking leads to a breakdown of confinement along the $\varphi_{-}$ direction. The reason is that the non-reciprocal terms change the acceleration function, such that acceleration and deceleration are not cancelled anymore, as in the conserving case. One can see this visually in the left plot in Fig.~\ref{fig: eom orbit}, wherein along the strip decelerating portions clearly dominate $\ddot{\varphi}_{-}$, hence the angular frequency increases constantly. The growth rate can presumably be related to a quantity like the integral over the strip.  In the oscillatory case, such as the middle plot, the confinement in both directions is preserved, although the non-reciprocal terms lead to a bending of the rectangle. In both plots the initial condition lies close to the stable fixed point of $\ddot{\varphi}_{+}$, with the stable / unstable fixed point of $\varphi_{-}$ leading to oscillatory / growth behavior respectively. The initial condition in the last plot is close to the unstable fixed point of $\ddot{\varphi}_{+}$ and the stable one for $\ddot{\varphi}_{-}$; consequently the dynamics is confined along $\varphi_{-}$ while traversing the full $\varphi_{+}$ range. The energy growth is small though, presumably due to the still near cancellation of acceleration and deceleration along the $\varphi_{+}$ direction.

The fixed points also explain the regions of large energy growth in the main plot in Fig.~\ref{fig: ear overview}. The region with high growth form exactly around those fixed points, with the stable fixed point in $\ddot{\varphi}_{+}$ leading to a rather well-defined region, while the unstable one is surrounded by more intricate dependencies on the initial conditions. 

In Fig.~\ref{fig: ear diagram} we show similar plots for various values of $qa$, shifting the axes, such that the stable fixed point (in both directions) is at the center and the unstable ones form the boundary. In the figure one observes that the region of large growth increases in size initially, but shrinks subsequently at larger values of $qa$. This can be explained by realizing that $qa \to \infty$ corresponds to a perfect conductor, which as discussed in Appendix~\ref{app: perfect cond} has reciprocal couplings.

 \subsection{\label{sec: long time dynamics} Long-time dynamics}
\begin{figure}[htbp]
 \centering
 \includegraphics[width=0.97\linewidth]{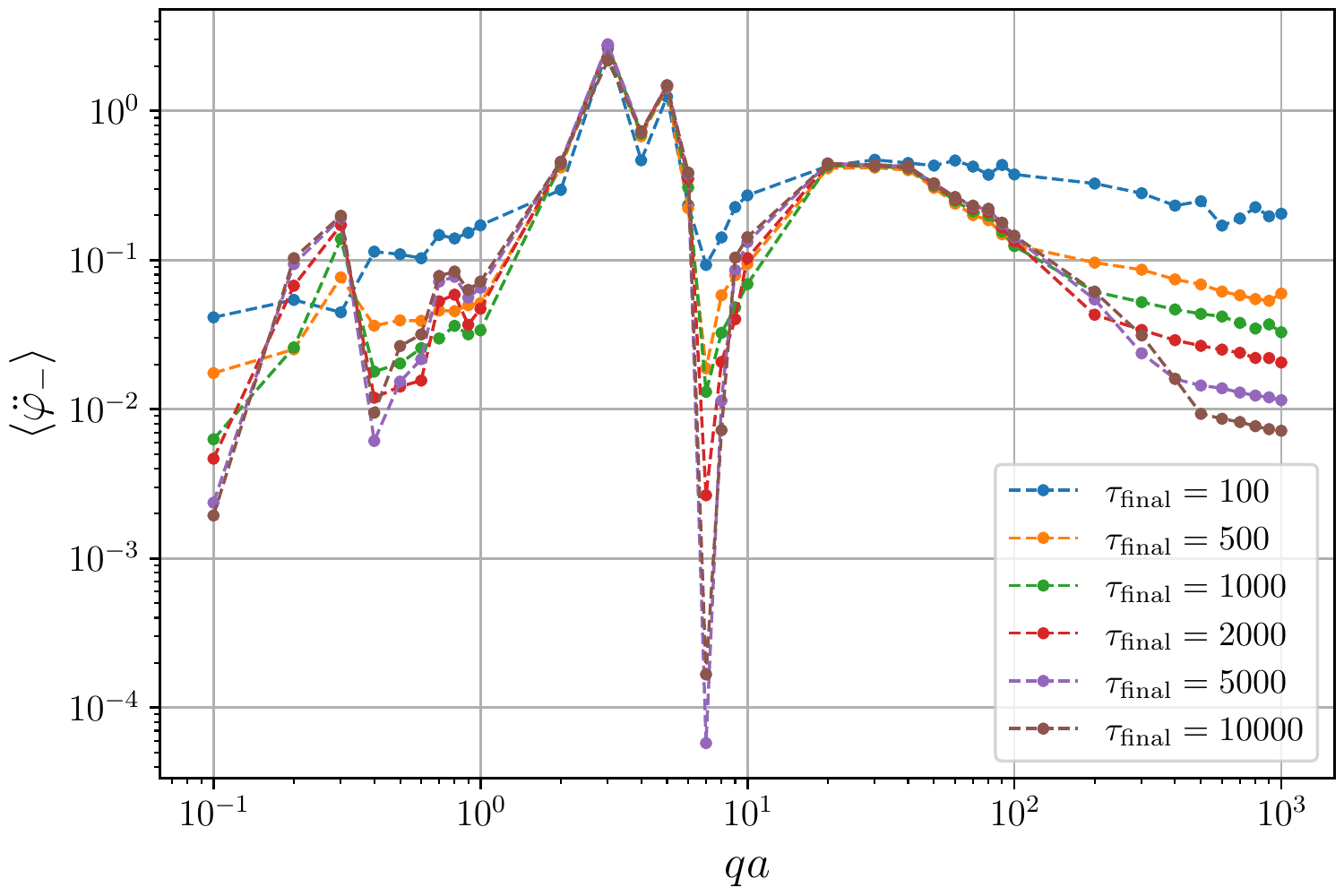} 
 \caption{\label{fig: long-time vary qa} Long-time behavior of the average acceleration rate with $\theta=45^{\circ}$, $z_{0} / a=0.1$, and initial condition $\varphi_{+}=3\pi/2, \; \varphi_{-}=0$. Different simulation times $\tau_{\mathrm{final}}$ are plotted. We observe three different types of behavior: decay with $\tau_{\mathrm{final}}$ mostly at the edges, convergence with $\tau_{\mathrm{final}}$ around the middle and a convergence followed by a decay at isolated points. Decay can be explained by oscillatory motion / confined motion in phase space, while the convergence indicates motion corresponding to an energy growth.}
 \end{figure}

 The results presented in the previous section were based on simulation times of $\tau_{\mathrm{final}}=20$ or $\tau_{\mathrm{final}}=100$. We have seen in Fig.~\ref{fig: eom orbit}, that this time is already enough to explore an extended region in phase space thoroughly. Yet, the question remains, whether the dynamics ultimately leaks into further parts of phase space. Trying to answer this, we analyze the dynamics for much larger times of up to $\tau_{\mathrm{final}}=10000$, hence several thousands of "cycles" given that the oscillation frequency of $\varphi_{+}$ is of order one (see for instance Fig.~\ref{fig: dipole dynamics}). In Fig.~\ref{fig: long-time vary qa} we plot the resulting average acceleration for a single initial condition ($\varphi_{+}=3\pi/2, \; \varphi_{-}=0$), but varying simulation times and values of $qa$. In the figure one can (roughly) identify converging and decaying behaviors of the rate.
 From earlier observations, we recognize that the decay corresponds to oscillatory behavior, while the convergence occurs due to growth as discussed in Sec.~\ref{sec: dynamics}. Though it is unclear what would happen in a case, wherein the entire phase space is explored. In any case, we see that confinement at some parameter values persists even to very long times. Whether it remains up to infinite times, can of course not be answered by numerical simulations definitely. However, in reality one does not expect these times to be practical anyway due to damping, which is discussed next.

\subsection{\label{sec: dyn with damping} Dynamics with damping}

 Finally, we consider the dynamics including damping. Simulations for various parameter values and values for the (dimensionless) damping coefficient $\tilde{\eta}=t_{\mathrm{sys}}/t_{\mathrm{damp}}$ suggest, that with damping a steady state is reached by the dynamics. For strong damping this steady state is essentially a rest state, but for values of $\tilde{\eta} \approx 1$ a steady state with finite angular \emph{frequency} (modulo some oscillations) can be reached, with some energy being absorbed in the process. The magnitude of the damping is chosen based on a reasonable experimental setup, as outlined in Appendix~\ref{app: experiment}. In Fig.~\ref{fig: ear with damping}, we again plot the average acceleration as a function of the initial conditions for select values of $qa$. Three different simulation times are plotted, since with damping we expect the averaged acceleration to depend on the simulation time even in the case of initial energy growth. We again observe clear regions wherein energy is absorbed by the dipoles, however many of the more intricate features appear to be "washed out" by the damping. In fact, for $qa=10$ the entire diagram seems uniform across the initial conditions. The strong dependence on the simulation time, as seen in the colors, signifies that the timescale of reaching the steady state is comparable to the times depicted in the figure.
 \begin{figure*}[htbp]
 \centering
 \includegraphics[width=0.99\linewidth]{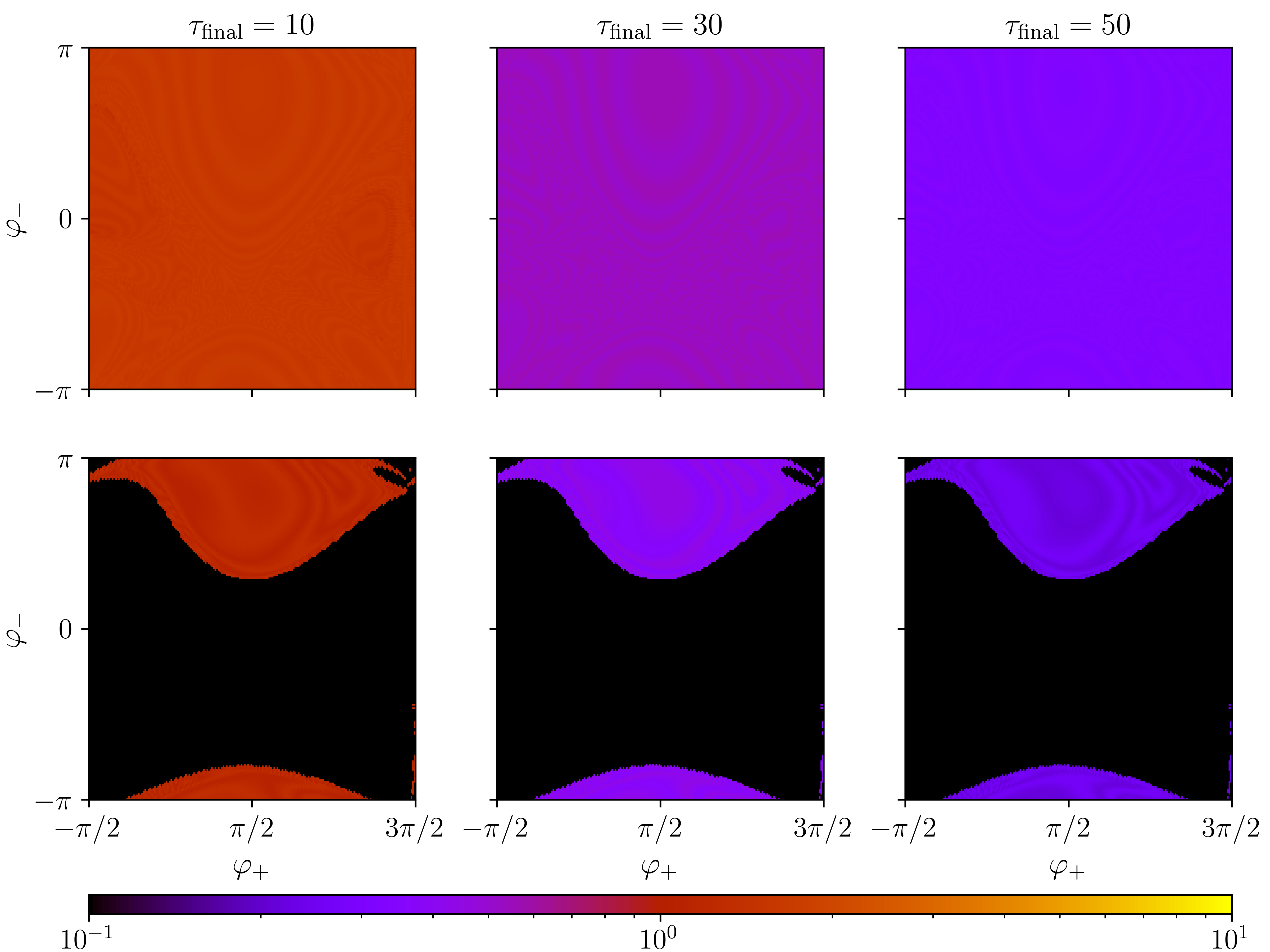}
 \caption{\label{fig: ear with damping} Average acceleration with damping ($\tilde{\eta}=1$) with $\theta=45^{\circ}$ and $z_{0} / a=0.1$ for $qa=10$ (upper) and $qa=100$ (lower) based on different simulation times. Even with damping there is still an energy absorption, even though it is strongly suppressed for $qa=100$. The average acceleration depends on the simulation time, indicating that the steady state is reached or nearby for the chosen times. Compared to the undamped case many of the more intricate dependencies on the initial conditions appear to be washed out.}
 \end{figure*}
 
\begin{figure}[htbp]
 \centering
 \includegraphics[width=0.97\linewidth]{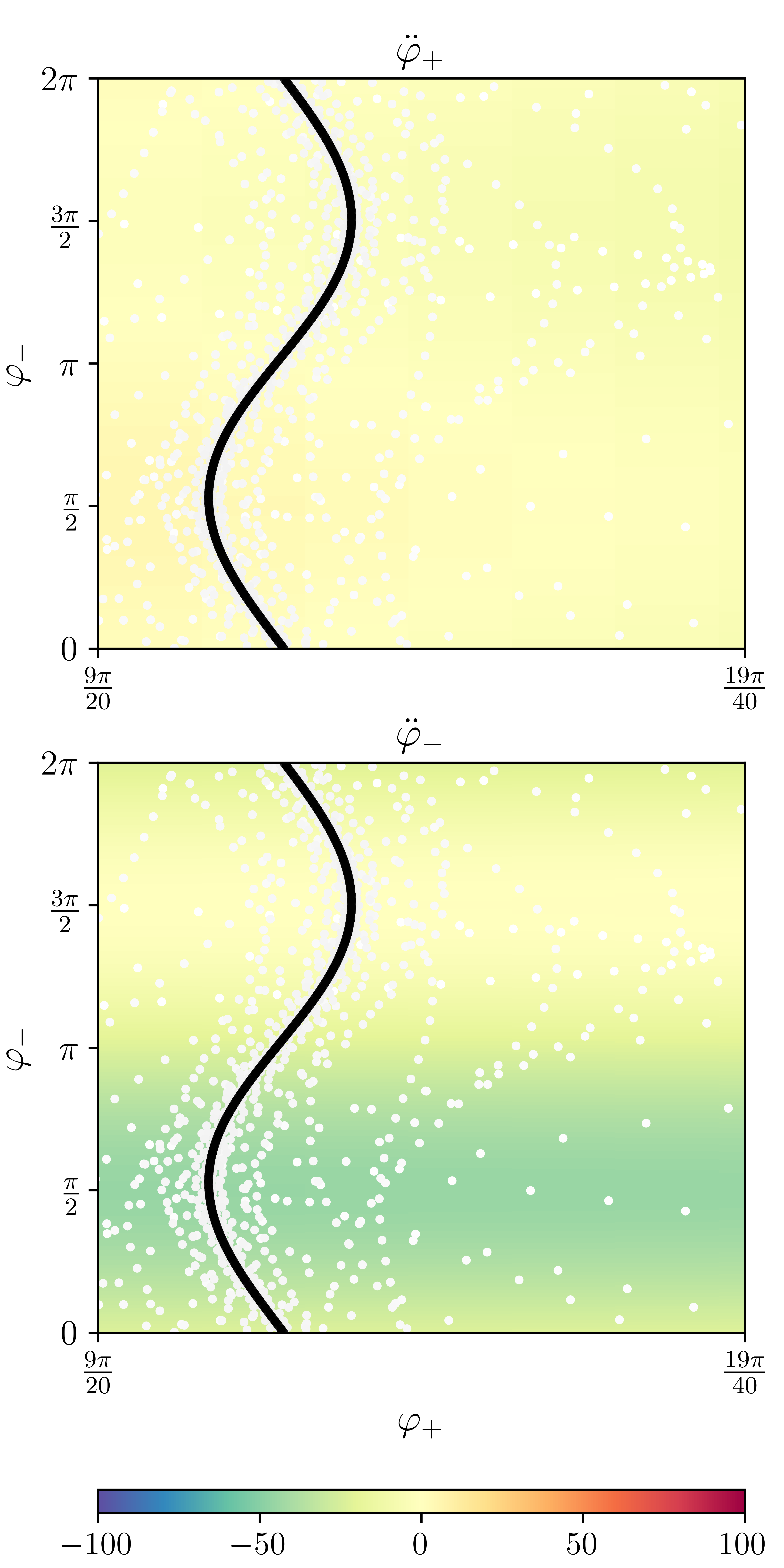}
 \caption{\label{fig: eom with damping}  Dynamics in parameter space with damping $\tilde{\eta}=1$ at $qa=10$, $\theta=45^{\circ}$, and $z_{0} / a=0.1$ starting from the initial condition $\varphi_{+}=\pi/2, \varphi_{-}=0$. The total evolution time is $\tau_{\mathrm{final}}=100$. The dynamics converges onto a curve with a very small extent in $\varphi_{+}$ (note the small range in $\varphi_{+}$ depicted in the plot). Visualizing the data in different ways suggests that a steady state is reached.}
 \end{figure}

 In phase space the steady state seems to correspond to motion on a curve, as can be seen in Fig.~\ref{fig: eom with damping}. The curve is strongly confined on the $\varphi_{+}$-axis, while being open along the $\varphi_{-}$-axis. However, visualizing the values for $\omega_{\pm}$ shows a steady state with some oscillations on top, therefore, deceleration and acceleration with respect to $\ddot{\varphi}_{-}$ should be balanced along the path, even though visually it is not clear how this balancing occurs exactly.

 \section{\label{sec: conclusion} Conclusion}
Overall, we have analyzed the dynamics of a pair of rigid XY dipoles with non-reciprocal interactions due to a moving conductor in detail, building on the work of Prat-Camps et al.~\cite{prat-camps_circumventing_2018}. We have shown, that the non-reciprocal terms have a significant effect on the dynamics leading to the possibility of energy absorption from the conductor. Here we found a very strong dependence on the initial conditions and provided a qualitative interpretation in terms of phase space confinement. These phenomena may be related to notions in dynamical systems such as chaos~\cite{strogatz_nonlinear_2019}. Furthermore, we have also argued, that many of these effects could well be observable in experiments, for example in setups as in~\cite{mellado_macroscopic_2012, concha_designing_2018, cisternas_stable_2021}. The results on statistical physics of non-reciprocal systems~\cite{ivlev_statistical_2015, fruchart_nonreciprocal_2021} motivate the experimental and theoretical study of the described system as well as possible future studies of many-body systems with non-reciprocal magnetic interactions.

The question remains though, of how the presented results might transfer to many-body systems. In particular, the strong dependence on initial conditions may lead to a "washing out" of the absorption effect, but the opposite could also be true. However, thinking about this question, one can make the observation that the number of degrees of freedom can be reduced by preparing the system in specific initial states on a lattice. Consider for example an initial state with all dipoles at rest and with the same orientation on a square lattice: the torques on all dipoles are equal, and thus the entire dynamics will be equal for any single dipole. Therefore, in this case, we would have an \emph{exact} mean-field description, which could be reduced to the degrees of freedom of a single dipole. Since the square lattice is a bipartite lattice, we could also get a reduction to a pair of dipoles, by preparing a state with equal orientations on each sub-lattice. As the couplings decay fast, the resulting model would most likely be close to the bare two-dipole case that we focused on throughout this article. Clearly, this idea can be generalized to an increasing number of dipoles, with an appropriate choice of the lattice. Therefore, we expect that the few-body effects can persist in the many-body case and thus their study can also be motivated from this perspective.

\begin{acknowledgments}
We are grateful to Patrick Maurer and Oriol Romero-Isart for stimulating discussions and collaboration in the first stage of the project. We thank Peter Derlet and Markus M\"uller for helpful discussions. We acknowledge support by the Austrian Science Fund FWF within the DK-ALM (W1259-N27). 

The computations and figures in this work have been obtained mostly using \textit{Python}~\cite{langtangen_primer_2009}, in particular with the (free and open) libraries \textit{Numba}~\cite{lam_numba_2015}, \textit{Numpy}~\cite{vanderwalt_numpy_2011, oliphant_guide_2015}, \textit{SciPy}~\cite{virtanen_scipy_2020} and \textit{Matplotlib}~\cite{hunter_matplotlib_2007},  and partially using \textit{Mathematica}~\cite{wolfram_mathematica_2020}. The data and code for this article is freely accessible at~\cite{rakcheev_dataset_2022}.
\end{acknowledgments}

\appendix
\section{\label{app: Derivation of the Potential Energy} Derivation of the coupling matrix}
In this section we derive the coupling matrix $\mathbb{K}$ of a coplanar configuration of magnetic dipoles based on the setup presented in Sec.~\ref{sec: setup} and Fig.~\ref{fig: sketch of geom}.  As discussed in Sec.~\ref{sec: couplings}, in principle only the field of a single dipole needs to be derived, however here we explicitly include all dipoles to emphasize that the total field is a superposition. Although we are interested in the non-relativistic limit $\beta\equiv v/c \ll 1$, where $c$ denotes the speed of light in vacuum, we first apply a fully relativistic treatment to the problem and perform all limits on the final expressions. 

\subsection{Electromagnetic field generated by the dipoles}
\label{ssec: Electromagnetic Field Emitted by the Lattice}
Our goal is to obtain the electromagnetic field generated by the dipoles in the rest frame of the conducting half-space, i.e. lab frame. The magnetization $\mathbf{M}'(\mathbf{r}')=\sum_i \mathbf{ m}'_i \delta(\mathbf{ r}'-\mathbf{r}'_i)$ in the lattice frame translates, via a Lorentz transformation, to a polarization $\mathbf{P}(\rr,t)=\sum_i \mathbf{p}_i \delta(\rr-\rr_i)$ and magnetization $\mathbf{M}(\rr,t)=\sum_i \mathbf{m}_i \delta(\rr-\rr_i)$ in the lab frame. Each dipole $i$ is characterized by its position
\begin{equation}\label{eqApp:position_vecotr}
    \mathbf{r}_i=(vt+x_i/\gamma)\uv_x+y_i\uv_y+z_0\uv_z,
\end{equation}
and its electric and magnetic moment
\begin{align}
\mathbf{p}_i=&(\beta/c)(-m_i^z \uv_y+m_i^y\uv_z),\\
\mathbf{m}_i=&(m_i^x/\gamma)\uv_x+m_i^y\uv_y+m_i^z\uv_z,
\end{align}
where $\gamma^{-1}=\sqrt{1-\beta^2}$ denotes the Lorentz factor~\cite{vekstein_electromagnetic_1997}. The polarization and magnetization give rise to a charge density $\rho(\rr,t)=-\nabla\cdot \mathbf{P}(\rr,t)$ and a current density $\mathbf{j}(\rr,t)=\partial_t \mathbf{P}(\rr,t)+\nabla\times \mathbf{M}(\rr,t)$. It immediately follows that the electric and magnetic field generated by dipole $i$ can, in the spectral domain, be expressed in terms of the dyadic Green's function $\mathbf{G}(\rr,\rr',\omega)$
\begin{align}
\label{eq: app1}
\mathbf{E}_i(\rr,\omega)=&\irm\mu_0\omega\rmint_{\mathbb{R}^3}\text{d}\rr'\mathbf{G}(\rr,\rr',\omega)\cdot \mathbf{j}_i(\rr',\omega),\\
\label{eq: app2}
\mathbf{B}_i(\rr,\omega)=&\mu_0\nabla\times\rmint_{\mathbb{R}^3}\text{d}\rr'\mathbf{G}(\rr,\rr',\omega)\cdot \mathbf{j}_i(\rr',\omega),
\end{align}
where $\mu_0$ denotes the vacuum permeability. Note that throughout this derivation we use the convention $\mathbf{f}(\rr,\omega)=\rmint_{\mathbb{R}}\text{d}t\, \mathbf{f}(\rr,t)\exp(\irm \omega t)$ for the Fourier Transform. The dyadic Green's function is the solution of the the inhomogenous Helmholtz equation
\begin{multline}
  \nabla\times\nabla\times \mathbf{G}(\rr,\rr',\omega)-(\omega/c)^2\epsilon(\rr,\omega)\mathbf{G}(\rr,\rr',\omega)\\
  =\delta(\rr-\rr')\mathbb{I},
\end{multline}
with the relative permittivity  $\epsilon(\rr,\omega)=1+[\epsilon(\omega)-1]\Theta(-z)$. We evaluate the electromagnetic fields in the upper half-space, i.e. $z>0$, where the Green's function can be subdivided into a bulk part and a scattering part  $\mathbf{G}(\rr,\rr',\omega)=\mathbf{G}_\text{b}(\rr,\rr',\omega)+\mathbf{G}_{\text{s}}(\rr,\rr',\omega)$ for $z,z'>0$. The bulk Green's function describes the evolution of electromagnetic fields in free space whereas the scattering Green's function describes the evolution of electromagnetic fields scattered by the conducting half-space. The total electromagnetic field reads 
\begin{align}
 \mathbf{E}(\rr,t)=&\sum_i[\mathbf{E}_i^\text{b}(\rr,t)+\mathbf{E}_i^\text{s}(\rr,t)],\\
 \mathbf{B}(\rr,t)=&\sum_i[\mathbf{B}_i^\text{b}(\rr,t)+\mathbf{B}_i^\text{s}(\rr,t)].
\end{align}
One can derive the bulk part of the electromagnetic field by either using the well known bulk Green's function for free space and calculate the field via Eq.~\eqref{eq: app1} and Eq.~\eqref{eq: app2} or by simply Lorentz transforming the static magnetic field created by the coplanar configuration of magnetic dipoles to the lab frame. Both approaches are straightforward and lead to the same well known expressions \cite{jackson_classical_1998}. The scattering Green's function for a half-space has a well known expression is also known \cite{buhmann_dispersion_2012} and using it one can obtain an expression for the coupling matrix. 
\subsection{Coupling matrix}
In the non-relativistic regime, where $\beta\ll 1$ and $\gamma \simeq 1$ the electric contribution is vanishingly small and can safely be neglected when compared to the magnetic contribution, and we arrive at the coupling matrix
\begin{equation*}
    \mathbb{K}_{\mathrm{dpl}}(\mathbf{r})=\frac{\mu_{0}}{4\pi}\frac{3\hat{\mathbf{r}} \otimes \hat{\mathbf{r}}-\mathbb{I}}{r^{3}}.
\end{equation*}
As expected, in the non-relativistic regime, this expression agrees with well known free-space dipole-dipole interaction term. 

For the scattering part, a lengthy but straightforward calculation shows that in the non-relativistic regime the induced part of the coupling matrix reads
\begin{equation}
\label{eq: coupling matrix full integral}
\mathbb{K}_{\mathrm{ind}}(\mathbf{r})=\frac{\mu_{0}}{8\pi^2 z^{3}_{0}}\rmint\limits_{0}^{\infty}\rmint\limits_{0}^{2\pi}\drm\xi \drm\phi \; \erm^{-2 \xi} \xi^{2} r_{s}(\xi, \phi)\erm^{\irm\frac{\xi}{z_{0}}\mathbf{r} \cdot \mathbf{e}_{\rho}}\mathbb{M}(\phi),
\end{equation}
with the magnetic Reynolds number $R=\mu_0 \sigma v z_0$, and the cylindrical unit vectors $\uv_\rho,\,\uv_\phi,\uv_z$. Before analyzing the expression in detail, we note that this derivation can easily be generalizing to the non-coplanar case ( $z_i \neq z_j$). In the general case, the term $\exp(-2\xi)$ that needs to be replaced by $\exp(z_{ij}\xi)$ with $z_{ij}\equiv (z_i+z_j)/2$.

Note that the integrand decays exponentially in $\xi$ due to the factor $\exp(-2\xi)$. The relative permittivity will therefore contribute appreciably in a region around the characteristic frequency of the system $\omega_c \equiv \gamma v/z_0$. For non-relativistic velocities up to $v\leq 10^3 \text{m}\text{s}^{-1}$ and distances down to $z_0\geq 10^{-9}\text{m}$ the characteristic frequency is $\omega_c \leq 10^{12} \text{s}^{-1}$. Typical relaxation times for metals are on the order of $\tau \simeq 10^{-14}\text{s}$ \cite{ashcroft_solid_1976} which leads to $\omega_c \tau \leq 10^{-2}$.  Therefore, we can safely approximate the relative permittivity by $\epsilon(\omega)\simeq 1+\irm \sigma/(\epsilon_0 \omega)$.
\section{\label{app: coupling details} Details on couplings}
\subsection{Decomposition}
The expression for the coupling matrix can be decomposed into contributions with different parity upon (spatial) inversion and transposition. To decompose it we start by expanding the matrix in the integrand
\begin{align}
(\mathbf{e}_{\rho}+\irm\mathbf{e}_{z})(\mathbf{e}_{\rho}-\irm\mathbf{e}_{z})& \\ \nonumber
 &=\mathbb{M}_{+}(\phi)+\irm\mathbb{M}_{-}(\phi) \\ \nonumber
&=\begin{pmatrix}
\cos^{2}(\phi) & \cos(\phi)\sin(\phi) & 0 \\ 
\cos(\phi)\sin(\phi) & \sin^{2}(\phi) & 0 \\ 
0 & 0 & 1
\end{pmatrix} \\
&\;\; + \irm\begin{pmatrix}
0 & 0 & -\cos(\phi) \\ 
0 & 0 & -\sin(\phi) \\ 
\cos(\phi) & \sin(\phi) & 0
\end{pmatrix},
\end{align}
with the even and odd matrices under transposition $\mathbb{M}_{\pm}$ also satisfying $\mathbb{M}_{+}(\phi+\pi)=\mathbb{M}_{+}(\phi)$ and $\mathbb{M}_{-}(\phi+\pi)=-\mathbb{M}_{-}(\phi).$ Using this and that $\mathbf{e}_{\rho}(\phi+\pi)=-\mathbf{e}_{\rho}(\phi)$ and $r_{s}(\xi, \phi +\pi) = r^{*}_{s}(\xi, \phi)$, we can reduce the angular integral from $0$ to $2 \pi$ to an integral from $0$ to $\pi$
\begin{align}
& \rmint\limits_{0}^{2\pi}\drm\xi \drm\phi \; r_{s}(\xi, \phi)\erm^{\irm\frac{\xi}{z_{0}}\mathbf{r} \cdot \mathbf{e}_{\rho}}(\mathbf{e}_{\rho}+\irm\mathbf{e}_{z})(\mathbf{e}_{\rho}-\irm\mathbf{e}_{z}) \nonumber \\
=& \; 2\mathrm{Re}(r_{s})\bigg[\cos(\frac{\xi}{z_{0}}\mathbf{r} \cdot \mathbf{e}_{\rho})\mathbb{M}_{+}-\sin(\frac{\xi}{z_{0}}\mathbf{r} \cdot \mathbf{e}_{\rho})\mathbb{M}_{-} \bigg]\nonumber\\
&-2\mathrm{Im}(r_{s})\bigg[\cos(\frac{\xi}{z_{0}}\mathbf{r} \cdot \mathbf{e}_{\rho})\mathbb{M}_{-}+\sin(\frac{\xi}{z_{0}}\mathbf{r} \cdot \mathbf{e}_{\rho})\mathbb{M}_{+} \bigg].
\end{align}
We can now decompose this further into an even and an odd part under inversion
\begin{align}
&\mathbb{K}_{+}(\mathbf{r}) = \frac{\mu_{0}}{4 \pi^2 z^{3}_{0}}\rmint\limits_{0}^{\infty}\drm\xi \; \erm^{-2 \xi} \xi^{2}\rmint\limits_{0}^{\pi} \drm\phi \nonumber \\ 
 &\times \left[ \mathrm{Re}(r_{s})\cos(\frac{\xi}{z_{0}}\mathbf{r} \cdot \mathbf{e}_{\rho})\mathbb{M}_{+}- \mathrm{Im}(r_{s})\cos(\frac{\xi}{z_{0}}\mathbf{r} \cdot \mathbf{e}_{\rho})\mathbb{M}_{-}\right] \nonumber \\
&\mathbb{K}_{-}(\mathbf{r}) = -\frac{\mu_{0}}{4 \pi^2 z^{3}_{0}}\rmint\limits_{0}^{\infty}\drm\xi \; \erm^{-2 \xi} \xi^{2} \rmint\limits_{0}^{\pi} \drm\phi \nonumber \\  
&\times  \left[\mathrm{Re}(r_{s})\sin(\frac{\xi}{z_{0}}\mathbf{r} \cdot \mathbf{e}_{\rho})\mathbb{M}_{-} + \mathrm{Im}(r_{s})\sin(\frac{\xi}{z_{0}}\mathbf{r} \cdot \mathbf{e}_{\rho})\mathbb{M}_{+}\right].
\end{align}
As we can see, each combination off inversion and transposition symmetry is represented in the expressions.
Finally, the single-ion anisotropy is the limit $\mathbb{A}=\lim\limits_{\mathbf{r} \to 0}\mathbb{K}_{+}(\mathbf{r})$
\begin{equation}
\mathbb{A}=\frac{\mu_{0}}{4 \pi^2 z^{3}_{0}}\rmint\limits_{0}^{\infty}\drm\xi \; \erm^{-2 \xi} \xi^{2} \rmint\limits_{0}^{\pi} \drm\phi \;  \left[\mathrm{Re}(r_{s})\mathbb{M}_{+}-\mathrm{Im}(r_{s})\mathbb{M}_{-} \right],
\end{equation}
where one can check that by symmetry only the diagonal components $A_{xx}, A_{yy}, A_{zz}$ are non-vanishing.
\subsection{Reflection symmetries}
Analyzing the trigonometric functions in the integral expression of the couplings, one can show that the spatial dependence of individual couplings $K^{\pm}_{ab}$ does not only have an inversion symmetry, but even a quadrant symmetry with respect to $\theta$. A convenient way to derive the appropriate symmetries is to consider reflections of $\mathbf{r}$ along the $x$- or $y$-axis, described by the reflection matrices $\mathbb{R}_{x/y}$. We do not present the relatively straightforward derivations here and state the results instead:
\begin{align*}
&K^{+}_{xx}(\mathbb{R}_{x/y}\mathbf{r})=K^{+}_{xx}(\mathbf{r}), \; K^{-}_{xx}(\mathbb{R}_{x/y}\mathbf{r})=-/+K^{-}_{xx}(\mathbf{r})\\
&K^{+}_{xy}(\mathbb{R}_{x/y}\mathbf{r})=- K^{+}_{xy}(\mathbf{r}), \; K^{-}_{xy}(\mathbb{R}_{x/y}\mathbf{r})=+/- K^{-}_{xy}(\mathbf{r})\\
&K^{+}_{xz}(\mathbb{R}_{x/y}\mathbf{r})= K^{+}_{xz}(\mathbf{r}), \; K^{-}_{xz}(\mathbb{R}_{x/y}\mathbf{r})=-/+ K^{-}_{xz}(\mathbf{r})\\
&K^{+}_{yy}(\mathbb{R}_{x/y}\mathbf{r})= K^{+}_{yy}(\mathbf{r}), \; K^{-}_{yy}(\mathbb{R}_{x/y}\mathbf{r})=-/+ K^{-}_{yy}(\mathbf{r})\\
&K^{+}_{yz}(\mathbb{R}_{x/y}\mathbf{r})=- K^{+}_{yz}(\mathbf{r}), \; K^{-}_{yz}(\mathbb{R}_{x/y}\mathbf{r})=-/+ K^{-}_{yz}(\mathbf{r})
\\
&K^{+}_{zz}(\mathbb{R}_{x/y}\mathbf{r})= K^{+}_{zz}(\mathbf{r}), \; K^{-}_{zz}(\mathbb{R}_{x/y}\mathbf{r})=+/- K^{-}_{zz}(\mathbf{r}).
\end{align*} 
Note that for $K^{\pm}_{ba}$ the relationship is the same, since the transpose symmetry acts on both sides of the equation in the same way. These allow to reduce the computations of the couplings by a factor of four, but can also be used to test the numerical integration, which we did during the process.
\subsection{Transformation}
To bring the expressions into a form suitable for numerical integration by quadrature, we use the transformations $x=2 \xi / R$ and $u=\cos(\phi)$ (from $0$ to $\pi$ $\sin(\phi)=\sqrt{1-u^{2}}$) to rewrite the matrices as
\begin{widetext}
\begin{align}
&\mathbb{K}_{+}(\mathbf{r}) = \frac{\mu_{0}q^{3}}{32 \pi^2} \rmint\limits_{0}^{\infty}\drm x \; \erm^{-Rx} x^{2} \rmint\limits_{-1}^{1} \frac{\drm u}{\sqrt{1-u^2}} \; \cos \left[\frac{qrx}{2}\Omega(\theta, u)\right] \left[ \mathrm{Re}(r_{s})\mathbb{M}_{+} - \mathrm{Im}(r_{s})\mathbb{M}_{-}\right] \nonumber\\
&\mathbb{K}_{-}(\mathbf{r}) = -\frac{\mu_{0}q^{3}}{32 \pi^2}\rmint\limits_{0}^{\infty}\drm x \; \erm^{-Rx} x^{2} \rmint\limits_{-1}^{1} \frac{\drm u}{\sqrt{1-u^2}} \; \sin \left[\frac{qrx}{2}\Omega(\theta, u)\right] \left[\mathrm{Re}(r_{s})\mathbb{M}_{-} + \mathrm{Im}(r_{s})\mathbb{M}_{+}\right],
\label{eq: app couplings units}
\end{align}
\end{widetext}
wherein the "frequency" 
\begin{equation*}
    \Omega(\theta, u) = u\cos(\theta) + \sqrt{1-u^{2}}\sin(\theta),
\end{equation*}
the scattering function
\begin{equation}
r_{s}(x, u)=\frac{x-\sqrt{x^{2}-2\irm ux}}{x+\sqrt{x^{2}-2\irm ux}}
\end{equation} and
\begin{align}
M_{+}&=\begin{pmatrix}
u^{2} & u\sqrt{1-u^{2}} & 0 \\ 
u\sqrt{1-u^{2}}  & (1-u^{2}) & 0 \\ 
0 & 0 & 1
\end{pmatrix} \nonumber\\
 M_{-}&=\begin{pmatrix}
0 & 0 & -u \\ 
0  & 0 & -\sqrt{1-u^{2}} \\ 
u & \sqrt{1-u^{2}} & 0
\end{pmatrix}. 
\end{align}

\section{\label{app: perfect cond} Perfect conductor}
In case of a perfect conductor ($\sigma \to \infty$), the scattering function $r_{s}=-1$ is purely real~\cite{buhmann_dispersion_2012}, and the couplings can be calculated exactly (for instance using \textit{Mathematica}). The couplings obtained in this way are reciprocal. To evaluate the integrals analytically we first start with the integral over $x$
\begin{equation}
    q^{3}\rmint\limits_{0}^{\infty}(-1)\erm^{-Rx}x^{2} \exp \left[ \irm \frac{qr}{2}\omega x \right] \; \drm x = -\frac{2}{\left(z^{3}_{0}-i\frac{r}{2}\omega \right)^{3}},
   \end{equation}
whose real and imaginary parts correspond to the $\cos$ and $\sin$ integrals. Using this, the angular integrals can be evaluated to
\begin{align*}
K_{xx} &=  \frac{4 \pi  \left(3 r^2 \cos (2 \theta )+r^2-8 z_{0}^2\right)}{\left(r^2+4 z_{0}^2\right)^{5/2}}\\
K_{yy} &=  \frac{4 \pi  \left(-3 r^2 \cos (2 \theta )+r^2-8 z_{0}^2\right)}{\left(r^2+4 z_{0}^2\right)^{5/2}}\\
K_{zz} &=   \frac{8 \pi  \left(r^2-8 z_{0}^2\right)}{\left(r^2+4 z_{0}^2\right)^{5/2}}\\
K_{xy} &= \frac{12 \pi  r^2 \sin (2 \theta )}{\left(r^2+4 z_{0}^2\right)^{5/2}}\\
K_{xz} &=  \frac{48 \pi  r z_{0} \sin (\theta )}{\left(r^2+4 z_{0}^2\right)^{5/2}}\\
K_{yz} &=  \frac{48 \pi  r z_{0} \cos (\theta )}{\left(r^2+4 z_{0}^2\right)^{5/2}}.
\end{align*}
We can identify the length scale $\ell = \sqrt{r^{2}+4z_{0}^{2}}$, however at the moment it lacks a physical interpretation. The single-ion anisotropy can be obtained by taking the limit $r \to 0$ and gives
\begin{equation}
A_{xx}= A_{yy}=\frac{A_{zz}}{2} = -\frac{4 \pi}{z^{3}_{0}}.
\end{equation}
The $x$- and $y$-components are equal and half as large as the $z$-component. Therefore, in a Hamiltonian description we expect an effective XY-model. The reason for this is, that due to the negative sign, $z$-components of the magnetic moments have a higher energy cost than $xy$-components. An analysis of the relevant terms shows, that the angular dependence of the induced couplings is similar to the dipole-dipole one. Therefore, we expect only a slight change in the physics and do not analyze this setup in detail.

\section{\label{app: eom} Derivation of the equations of motion}
\subsection{Derivation from torque}
Using the couplings defined above we can expand the rhs
\begin{widetext}
\begin{align}
&\left(\mathbf{m}_{n}\times \mathbf{B}_{n}\right)\cdot \mathbf{e}_{z} = \sum\limits_{k \neq n}\left(\mathbf{m}_{n}\times \mathbf{B}_{k}\right)\cdot \mathbf{e}_{z} =\sum\limits_{k \neq n}\left(\mathbf{m}_{n}\times \mathbb{K}(\mathbf{r}_{kn})\mathbf{m}_{k}\right)\cdot \mathbf{e}_{z} \nonumber\\
=& m^{2}\sum\limits_{k \neq n} K_{xy}(\mathbf{r}_{kn})\cos(\varphi_{k}+\varphi_{n})+K_{yy}(\mathbf{r}_{kn})\sin(\varphi_{k})\cos(\varphi_{n})-K_{xx}(\mathbf{r}_{kn})\cos(\varphi_{k})\sin(\varphi_{n}).
\label{eq: eom from torque}
\end{align}
\end{widetext}
In the derivation $m$ is the norm of the dipole moment $|\mathbf{m}_{n}|=m \; \forall n$, and we use that $K_{xy}=K_{yx}$.
\subsection{Alternative derivation}
For reciprocal systems we can derive the equation from the Euler-Lagrange equations or Hamilton's equation. Throughout the derivation we assume that the dipole moments are confined to the $xy$-plane. The kinetic energy is given by
\begin{equation}
T=\frac{I}{2}(\omega^{2}_{1}+\omega^{2}_{2})
\end{equation}
and the potential energy by
\begin{equation}
V=-\mathbf{m}_{2}\mathbf{B}_{1}=-\mathbf{m}_{2}\mathbb{K}(\mathbf{r})\mathbf{m}_{1},
\end{equation}
where $\mathbf{r}$ is the displacement from $\mathbf{m}_{1}$ to $\mathbf{m}_{2}$. Expanding the expression we get
\begin{align*}
V =& -\mathbf{m}_{2}\mathbb{K}(\mathbf{r})\mathbf{m}_{1} \\
=& -m^{2}[ K_{xx}\cos(\varphi_{1})\cos(\varphi_{2}) + K_{yy}\sin(\varphi_{1})\sin(\varphi_{2})\\
&+K_{xy}\sin(\varphi_{1}+\varphi_{2})].
\end{align*}

The Euler-Lagrange equations for $\mathcal{L}=T-V$ are
\begin{equation}
\frac{\partial \mathcal{L}}{\partial q_{i}}=\frac{d}{dt}\frac{\partial \mathcal{L}}{\partial \dot{q}_{i}}.
\end{equation}
In our case $q_{i} \equiv \varphi_{i}$ and $\dot{q}_{i} \equiv \omega_{i}$. The rhs reads
\begin{equation}
\frac{d}{dt}\frac{\partial \mathcal{L}}{\partial \dot{q}_{i}}=I\dot{\omega}_{i}=I \ddot{\varphi}_{i}
\end{equation}
and the lhs matches the EOM derived from torque (without damping). In the non-reciprocal case the formalism can still be used, to obtain the EOM of one dipole in the field of the other. In this way Eq.~\eqref{eq: eom from torque} will be obtained for each individual dipole.
\subsection{Sum and difference variables}
For a pair of dipoles, it can be advantageous to consider the dynamics using the variables $\varphi_{\pm}=\varphi_{1} \pm \varphi_{2}$ and the corresponding angular velocities. Rewriting the EOM~\eqref{eq: eom from torque} in these variables yields
\begin{widetext}
\begin{align}
\ddot{\varphi}_{+} =& K^{+}_{xy}(\mathbf{r})\cos(\varphi_{+})+\frac{K^{+}_{yy}(\mathbf{r})-K^{+}_{xx}(\mathbf{r})}{2}\sin(\varphi_{+})+\frac{K^{-}_{yy}(\mathbf{r})+K^{-}_{xx}(\mathbf{r})}{2}\sin(\varphi_{-}) \nonumber\\
\ddot{\varphi}_{-} =& -K^{-}_{xy}(\mathbf{r})\cos(\varphi_{+})+\frac{K^{-}_{xx}(\mathbf{r})-K^{-}_{yy}(\mathbf{r})}{2}\sin(\varphi_{+})-\frac{K^{+}_{yy}(\mathbf{r})+K^{+}_{xx}(\mathbf{r})}{2}\sin(\varphi_{-}),
\end{align}
\end{widetext}
where the upper signs again denote the even and odd part upon inversion. We can see that for general reciprocal couplings (the transpose is symmetric in the $x/y$-couplings subspace) these equations decouple, and we expect oscillatory solutions, in the sense that the kinetic energy is bounded by energy conservation.

\section{\label{app: dyn with sia} Effects of the single-ion anisotropy}
\begin{figure*}[htbp]
\centering
\includegraphics[width=0.99\linewidth]{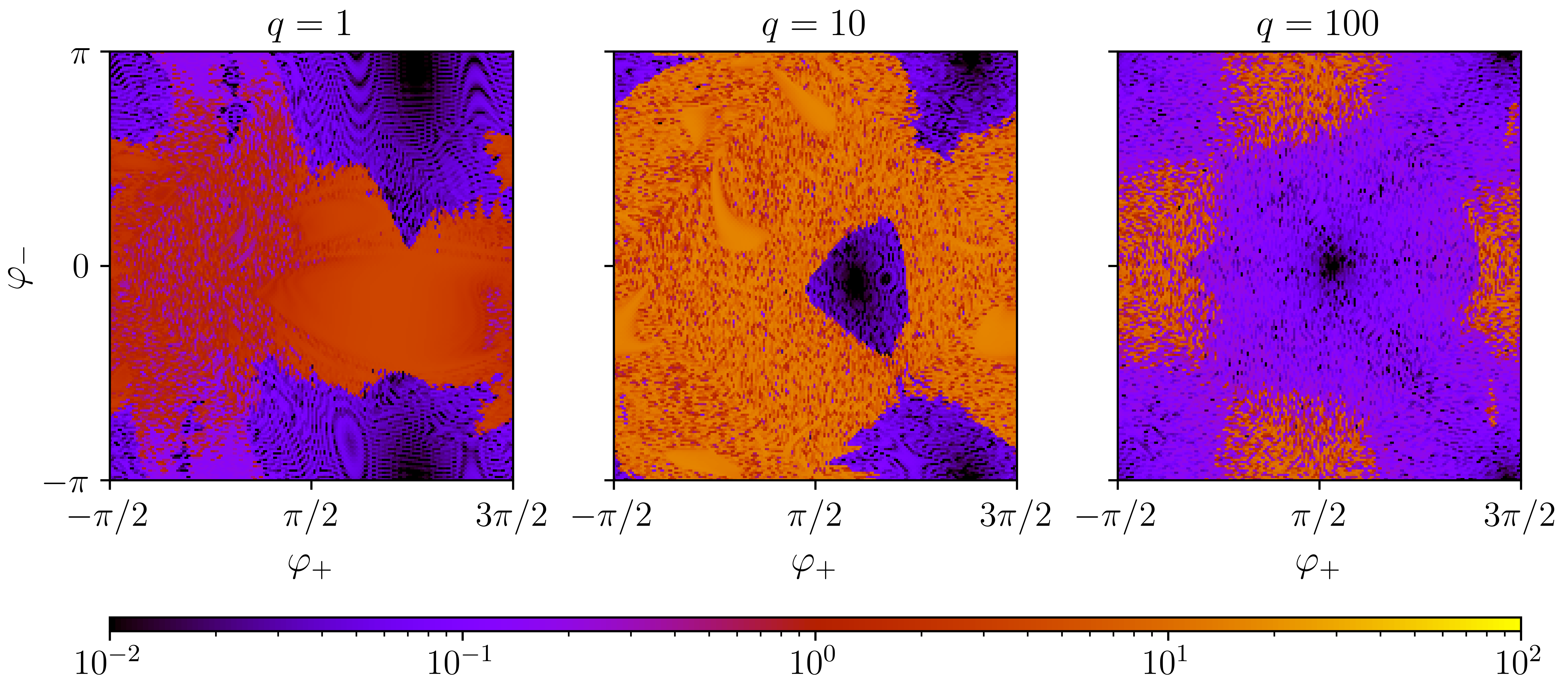}
\caption{\label{fig: ear with sia} Average acceleration rate with single-ion anisotropy for an angle of $45^{\circ}$, $z_{0}/a=0.1$, and three different values of $qa$ based on a simulation time of $\tau_{\mathrm{final}}=100$. We observe regions with significant absorption, even though the shape of these regions is significantly changed compared to the case without a single-ion anisotropy.}
\end{figure*}

In the simulations thus far, we have neglected the effects of the single-ion anisotropy. 
The reason is, that while it is formally easy to include it in the EOM, by simply adding the field, physically the situation is not so clear. In fact to create a torque, a rigid body has to be acted on the side, while formally the single-ion anisotropy act at the center of the dipole. This is hidden somewhat in the derivations by the implied assumption that the field from other dipoles does not vary too much across the physical extent of the dipole. In fact a discrepancy can be seen by comparing the EOM contribution of the field $\mathbb{A}\mathbf{m}$ in the torque derivation or the Lagrangian derivation. The Lagrangian derivation leads to a factor of $2$, due to square terms like $\sin^{2}(\varphi_{1})$ instead of $\sin(\varphi_{1})\sin(\varphi_{2})$. Intuitively, the Lagrangian derivation is more reliable, and since the effect of the new terms is stronger we use this in the simulations. Since the single-ion anisotropy is a reciprocal interaction, we expect it to favor a confined motion and to suppress energy absorption. Nevertheless, energy growth can be observed, as seen in Fig.~\ref{fig: ear with sia}, even though the shapes of the absorbing regions are changed significantly.
\section{\label{app: experiment} Parameter values}
In the experiments~\cite{mellado_macroscopic_2012, concha_designing_2018, cisternas_stable_2021} cylindrical magnetic Neodymium rods constrained to move in the $xy$-plane act as magnetic dipoles. The parameters vary across the cited works, with the ones in~\cite{mellado_macroscopic_2012} seeming most favorable for our setup, therefore we will use these here. The length of a rod is $\ell \approx 1.9 \times 10^{-2}\mathrm{m}$, the diameter $d \approx 1.5 \times 10^{-3} \;\mathrm{m}$, the mass $M \approx 0.28 \times 10^{-2} \;\mathrm{kg}$ and the saturation magnetization $M_{\mathrm{sat}} \approx 1.2 \times 10^{6} \;\mathrm{Am^{-1}}$. From these one can derive moment of inertia $I=\frac{1}{12}M\ell^{2} \approx 8.4\times 10^{-9} \; \mathrm{kg m^{2}}$, the  "magnetic charge" $Q=\pi \left(\frac{d}{2}\right)^{2}M_{\mathrm{sat}} \approx 2.03 \; \mathrm{Am}$ and subsequently the magnetic moment $m=Q\ell \approx 3.9 \times 10^{-2} \; \mathrm{Am^{2}}$.

Given these values, the timescale $t_{\mathrm{sys}}=\sqrt{\frac{32 \pi^{2} I a^{3}}{\mu_{0}m^{2}}}$ introduced in Sec.~\ref{sec: dynamics} can be estimated to
\begin{equation*}
    t_{\mathrm{sys}}\approx 38 a^{\frac{3}{2}} \; \mathrm{s},
\end{equation*}
where $a$ is the distance between dipoles measured in meters. The damping timescale $t_{\mathrm{damp}}=I/\eta$ is approximately $1s$. In dimensionless units the damping coefficient is 
\begin{equation*}
    \tilde{\eta}=t_{\mathrm{sys}}/t_{\mathrm{damp}} \approx 38 a^{\frac{3}{2}}.
\end{equation*}

As described in the main text, values for $qa$ where the non-reciprocity can have substantial effects lie within the range of $10^{-1}-10^{3}$ for the investigated distance to the plate $z_{0}/a=0.1$. For the most conductive non-magnetic metals (relative permeability $\mu \approx 1$) such as copper, aluminum, gold and silver, typical values for the conductivity (at room temperature) are around $\sigma \approx 4-6 \cdot 10^{7}\mathrm{Sm^{-1}}$, from which we get $qa = \mu_{0}\sigma va \approx 50-75va \;  \mathrm{sm^{-2}}$. 

Let us conclude with providing some estimates for experimental parameters needed to realize the values for the damping constant and other parameters used in Sec.~\ref{sec: dyn with damping}. Assuming that the experimental values for the magnetic moment, the moment of inertia and damping are set, one can adjust the damping by setting the distance $a$ to $\approx 1 \; \mathrm{dm}$. As a consequence, the velocity should be on the order of $1 \mathrm{ms^{-1}}$ to reach reasonable values of $qa$. The choice of $a$ implies that $z_{0} \approx 1 \; \mathrm{cm}$, which might pose an experimental challenge, but seems still realistic overall. Furthermore, near-field effects could arise from both the distance to the plate and the distance of the dipoles to each other, however given that $a$ is a multiple of $\ell$ and $z_{0}$ a multiple of $d$, these might not have a too large influence, but this would need to be checked for a concrete setup.

\bibliography{bib_dipoles}
\end{document}